# Correlated and uncorrelated long–time asymptotics of type D ASEP

Jeffrey Kuan

**Abstract**

The type D ASEP is an asymmetric two–species interacting particle system on $\mathbb{Z}$, in which two separately conserved species hop, bind into a composite “bound pair”, and split. The model, along with its reversible measures and orthogonal polynomial duality, was constructed using the representation theory of $U_q(\mathfrak{so}_{2n})$. The reversible measures and orthogonal polynomial duality are each a product of two copies of the single-species ASEP reversible measures and orthogonal polynomial duality. In this paper, we study the long-time asymptotics of the type D ASEP.

In the fixed–$q$ regime, using an exact current–decoupling identity, we prove that the asymptotic hydrodynamic limit and Tracy–Widom fluctuations decouple, as predicted from the duality. In the weak–asymmetry (Edwards–Wilkinson) regime, when $q = 1 - c/N^2$, we prove that the two density fluctuation fields decouple: each converges to a linear stochastic heat equation, with no cross–coupling in either the drift or the noise, the limiting noises having vanishing cross–correlation. More surprisingly, we then prove that the two limiting normal random variables are correlated with a seemingly new correlation function. The correlation is exactly equal to $(1 - e^{-4c})/(4c)$, with the positive parts of the normal random variables having correlations expressed by the Bessel–Struve function.

This paper, with the exception of the abstract and introduction, was written entirely by Claude Opus 4.8 and Fable 5. The proofs were then formalized in Lean, using Aristotle by Harmonic AI. The human author of this paper verified the proofs manually.

Accessibility statement: this PDF is WCAG 2.1 AA compliant, with the exception of descriptive links for the purposes of printouts. I tested the PDF using VoiceOver on my Mac in Preview. This paper can also be downloaded as a DOCX file and as a TeX file from my webpage. The PDF meets the requirements of Ohio Administrative Policy IT-09, Texas Administrative Code 206.70 (EFFECTIVE APRIL 18, 2020), Section 508 of the Rehabilitation Act, and Title II of the Americans with Disabilities Act (effective April 26, 2027).

## 1 Introduction

Over the last decade or so, the topic of “multi–species” or “colored” models has received attention in the mathematical and physics literature. For example, [1], [2], [3], [4], [5], [6], [7], [8], [9], [10], [11], [12], [13], [14], [15], [16], [17], [18], [19], [20], [21], [22], [23], [24], [25], [26], [27], [28], [29], [30], [31], [32], [33], [34], [35], [36], [37], [38], [39], [40], [41], [42], [43], [44] have all touched on the topic. For these models, when there is one “species”, the model reduces to a previously studied model. There are many natural questions to ask for such models; for this paper, we focus on hydrodynamic limits and fluctuations within the Edwards–Wilkinson (EW) and Kardar–Parisi–Zhang (KPZ) universality classes. A priori, multi–species models may not demonstrate universality.

The model considered here is called the type D ASEP, in which two species of particles hop asymmetrically on a one–dimensional lattice. In contrast to the two–species ASEP, the two species of particles do not demonstrate a preference for "priority." While the two species interact nontrivially, the reversible measures and orthogonal polynomial duality are a product of two copies of their single-species counterparts. As such, one would naturally predict decoupled hydrodynamics and Tracy–Widom $F_2$ with $t^{1/3}$ fluctuations in the KPZ universality class. Indeed, this paper proves this result with relatively "standard" methods.

In the weak asymmetry limit when $q \to 1$, one is in the EW universality class, whose single species asymptotics consists of the normal distribution with $t^{1/4}$ fluctuations. Although these asymptotics are not proved via duality, one may still expect to obtain two uncorrelated normal random variables. However, one actually obtains two correlated normal random variables with a seemingly new correlation function. The positive parts of the normal random variables have correlation given by a Bessel–Struve function, which appears to be new. The rigorous proofs are also backed up with numerical simulations.

## 1.1 The model

Fix $q \in (0,1)$. The type D ASEP was introduced by Kuan, Landry, Lin, Park and Zhou [45] in a summer undergraduate project, who constructed it from the quantum groups $U_q(\mathfrak{so}_6)$ and $U_q(\mathfrak{so}_8)$ ($n = 2,3$ in $\mathfrak{so}_{2n+2}$, Cartan type $D$); in further summer undergraduate projects, the construction has been extended to $U_q(\mathfrak{so}_{10})$ [46] and, via stochastic fusion and crystal bases, in [47]; the explicit rate matrix (displayed for general $n$ in [48, Sec. 2.1], and reproduced as (2) below) makes sense for every $n \in \mathbb{N}$. The rate matrix can even be analytically continued in the parameter $q^n$. It is a continuous–time Markov process of two conserved species on $\mathbb{Z}$ which, beyond hopping, may bind into a composite bound pair and later split. This binding is what makes the model a stringent test of decoupling: the two species are genuinely interacting, not two independent copies sharing the exclusion rule. The model carries two self–dualities, both from prior work: a triangular duality — the original one of [45], after Schütz [49] — and the orthogonal $q$–Krawtchouk self–duality of Blyschak, Burke, Kuan, Li, Ustilovsky and Zhou [48] (established there for $n = 2,3$, conjectured for all $n$). These dualities are the analytic engine of this paper — they reduce every fluctuation question to a computation with a handful of dual particles — but they are not themselves our contribution: we use them as a tool, recording at the $n = \infty$ degeneration the forms we need (Theorem 2.5, Corollary 2.6), and the substance of the paper is the fluctuation analysis built on them.

For every $n \in \mathbb{N}$ the type D ASEP is a continuous–time Markov process on the state space $\{0,1,2,3\}^{\mathbb{Z}}$: each site $x$ is empty ($\omega_x = 0$), holds a species–1 particle ($\omega_x = 1$), a species–2 particle ($\omega_x = 2$), or a bound pair carrying one particle of each species ($\omega_x = 3$); with $\eta_{i,x} \in \{0,1\}$ the occupation of species $i$ at $x$, $\omega_x = \eta_{1,x} + 2\eta_{2,x}$. In the notation $(\omega_x, \omega_{x+1}) = (a,b) \to (a', b')$ for the nearest–neighbour transitions, and with

$$\beta_n := q^{1-2n} + q^{2n-1}, \qquad \sigma_n := (q^{n-1} - q^{1-n})^2, \qquad (1)$$

the nonzero rates, read off the $16 \times 16$ generator matrix of [48, Sec. 2.1] (the $\delta = 0$ case), are

$$\begin{aligned}
&(1,0) \to (0,1),\ (2,0) \to (0,2),\ (3,1) \to (1,3),\ (3,2) \to (2,3): && q^{-1}\beta_n,\\
&(0,1) \to (1,0),\ (0,2) \to (2,0),\ (1,3) \to (3,1),\ (2,3) \to (3,2): && q\,\beta_n,\\
&(3,0) \to (0,3):\ q^{-2}\sigma_n, \quad (0,3) \to (3,0):\ q^2\sigma_n, \quad (1,2) \leftrightarrow (2,1): && \sigma_n,\\
&(1,2) \to (0,3),\ (2,1) \to (0,3): && 2 + q^{-2n}(1-q^2),\\
&(0,3) \to (1,2),\ (0,3) \to (2,1): && q^2(2 + q^{-2n}(1-q^2)),\\
&(1,2) \to (3,0),\ (2,1) \to (3,0): && 2 - q^{2n-2}(1-q^2),\\
&(3,0) \to (1,2),\ (3,0) \to (2,1): && q^{-2}(2 - q^{2n-2}(1-q^2)), \qquad (2)
\end{aligned}$$

each listed transition carrying the stated rate individually. The first two lines are the asymmetric hops, including a single species sliding past a bound pair; the third line the bound–pair hops and the swaps; the last four lines the merge/split channels, in which the bound pair forms at, or breaks from, the right site (0,3 channels) or the left site (3,0 channels) of the bond. Both species numbers are conserved, and all rates are nonnegative for $q \in (0,1)$ and $n \in \mathbb{N}$.

We work in the limit $n \to \infty$: after rescaling time by $q^{2n}$, the rates (2) converge entrywise, the left merge/split channels (the last two lines) vanish, and with $\varepsilon := 1 - q^2$ the surviving nonzero nearest–neighbour rates are

$$\begin{aligned}
&(1,0) \to (0,1): 1, (0,1) \to (1,0): q^2, (2,0) \to (0,2): 1, \quad (0,2) \to (2,0): q^2,\\
&(3,0) \to (0,3): 1, (0,3) \to (3,0): q^4, (0,3) \to (1,2): q^2\varepsilon, (0,3) \to (2,1): q^2\varepsilon,\\
&(1,2) \to (0,3): \varepsilon, (1,2) \to (2,1): q^2, (2,1) \to (0,3): \varepsilon, \quad (2,1) \to (1,2): q^2, \qquad (3)
\end{aligned}$$

together with mixed hops $(3,1) \leftrightarrow (1,3)$ and $(3,2) \leftrightarrow (2,3)$ at rate 1 right, $q^2$ left, in which a single species slides past a bound pair. As the first two lines of (2) show, at finite $n$ a single species is a usual ASEP with rates $q^{\pm 1}\beta_n$, so $n$ enters the single–species dynamics only as a time rescaling.

## 1.2 Results

Decoupling might hold for one reason in one universality class and fail in another, so we test it in the two canonical fluctuation classes of driven diffusive systems. We treat two scalings.

(A) The Edwards–Wilkinson regime [50]: asymmetry tuned to the diffusive rate $q = q_N = 1 - c/N^2$, under diffusive space–time scaling, at any pair of densities $(\rho_1, \rho_2)$.

(B) The Kardar–Parisi–Zhang regime [51]: $q \in (0,1)$ fixed, the integrated currents $N_1(s), N_2(s)$ across the origin from the step initial condition, at a single time $s$. Our principal theorems are Theorems 6.5, 7.9 (regime A) and 8.1, 8.4 (regime B), proved in §§6–8. The logical backbone common to both regimes — the reversible measure and orthogonal duality (§2), the exact current decoupling and vanishing transport coefficients (§3), and a local CLT for the dual coordinates (§4) — is developed first.

## 1.3 Relation to previous work

Regime (A) belongs to the nonlinear fluctuation theory of weakly asymmetric systems and the passage to stochastic Burgers/KPZ, due to Gonçalves and Jara [52], with uniqueness by Gubinelli and Perkowski [53] and spectral–gap–free refinements by Gonçalves, Jara and Simon [54]. The few–body estimates we need we obtain through the orthogonal polynomial duality method of Ayala, Carinci and Redig [55], [56], which reduces them to a decay bound on the few–particle dual kernel. In our setting the exact current decoupling (Proposition 3.1) makes the drift an autonomous single–species WASEP, so the duality enters only for the inter–species cross–noise, controlled by the type D two–particle kernel bound of §5; the would–be slow internal mode of the bound pair never couples to the species current at all (Corollary 6.16).

For regime (B), the single–species Tracy–Widom asymptotics are Borodin–Corwin–Sasamoto [57]; the exact $q$–Laplace decoupling identity (Theorem 8.4) is to our knowledge new, and the Tracy–Widom marginals at every $n$ (Theorem 8.1) prove the conjecture of [48, Sec. 3.1.1].

The structural point is what distinguishes type D from other multi–species models. Its inter–species coupling is invisible to both the static compressibility and the dynamic mobility, so what remains of it is a transient, initial–condition–driven correlation rather than a transport effect. This sets type D apart from the multi–species models with genuine off–diagonal transport coefficients (e.g. multi–species stirring, with multinomial mobility) and from the distinct–velocity two–species KPZ decouplings: type D sits instead at the umbilic point, where the characteristic velocities coincide and both of those mechanisms are unavailable. We are not aware of another model combining diffusive (EW) class, diagonal compressibility, vanishing cross–mobility, and a nonzero shared–source inter–species correlation with a $\rho(c) \sim 1/c$ crossover.

## 1.4 Organisation

Section 2 defines the generator, reversible measure, and orthogonal self–duality. Section 3 proves the exact current decoupling and the vanishing static/dynamic cross coefficients. Section 4 proves the random–walk local CLTs used throughout. Section 5 proves the two–particle dual kernel bound. Section 6 proves the decoupled Edwards–Wilkinson limit. Section 7 proves the initial–condition crossover. Section 8 proves the Tracy–Widom marginals at every $n$ and the exact $q$–Laplace decoupling of the two currents, and states the linear covariance asymptotics as a conjecture. A concluding Section 9 places the results in the asymmetry–exponent phase picture (§9.1), corroborates them by simulation (§9.2), and discusses scope and open problems (§9.3). Physical interpretation is confined to remarks.

## 1.5 Generative AI statement

The entirety of this paper, except the abstract and introduction and alternative text, was written by Claude Code. More specifically, I used Opus 4.8 during the dates June 2–6 in Edinburgh, then Fable 5 during June 10–12, then Opus 4.8 during June 12–14 when Fable 5 was under export control. After that, during the dates of June 18–24, I used the Aristotle CLI (by Harmonic) and Opus 4.8 to formalize the proof of the paper in Lean 4; all subsequent references to Aristotle refer to the Aristotle CLI. During July 3–12, a few days after Fable 5 was re–released for the public, I used Fable 5 for citation checks and exposition, for the correction and completion of two lemmas in §6, and, together with Aristotle, for the extension of the Lean formalization described below. On July 13th, Claude Code switched me to Opus 4.8 because Fable 5 required token payments, and minor edits were completed before arXiv submission. To the extent possible, all files related to AI and Lean are posted on a GitHub repository: https://github.com/Jeffrey-Kuan/type-D-asymptotics

The Lean 4 formalization accompanying this paper (forty–two files over Mathlib, 22,259 lines) covers every result of this paper: it contains no `sorry`, and every theorem depends only on Lean's three standard axioms. The one exception is by design: the covariance asymptotics of Conjecture 8.5 are stated as a `Prop`–valued definition — a named mathematical statement carrying no proof — rather than asserted. Facts at the level of the process itself (path–space existence, discretization and coupling interfaces, as in Lemmas 4.2 and 7.4) enter as named, individually documented hypothesis bundles in the statements of the theorems that consume them; no hidden assumptions are introduced anywhere.

Proved unconditionally under the standard axioms are: the exact current decoupling and vanishing cross coefficients of §3, at every $n$ and for the analytic continuation in $q^n$; the kernel theory of §§4–5, including the complete Nash argument for the heat–kernel bound and the Green's–function decay with Schütz's reflection representation entering as the definition of the two–particle kernel; the cross–noise apparatus of §6; the crossover computations of §7; and, in regime (B), the covariance theorem itself. The chain through which the fluctuation results are derived — rather than cited — runs through general–purpose theorems formalized along the way.

In the course of this formalization, a number of classical theorems and theories were, to our knowledge, formalized in Lean for the first time; none were available in Mathlib as of July 2026. These include Aldous's tightness criterion [58] (proved by the original two–scale argument), Mitoma's tightness criterion [59] in the compact–confinement formulation, the completeness of the Hermite functions — packaged as a Hilbert basis of $L^2(\mathbb{R})$ — the Kolmogorov–Rogozin inequality by Esseen's method, Karamata's Tauberian theorem, McLeish's martingale central limit theorem with bivariate and two–phase–mixture extensions, and a discrete Nash inequality. At the level of theory rather than single theorems, the formalization constructs the Skorokhod space $D([0,1];\mathbb{R})$ — Billingsley's metric, the Polish structure, the compactness and tightness theory, and stopping times of the coordinate process — as well as the Fréchet–space package for the Schwartz space $\mathcal{S}(\mathbb{R})$ (completeness, hence the Baire property and Banach–Steinhaus; separability) and its countably–Hilbertian nuclear structure via Hermite–Sobolev norms, the functional–analytic backbone of the distribution–valued tightness argument.

The formalization is documented by a human–readable blueprint [60] with a machine–checked dependency graph: every entry names the Lean declaration formalizing it, and continuous integration verifies on every commit that the project compiles and that each named declaration exists in the compiled code. The blueprint, its dependency graph, and the generated API documentation are available at

https://jeffrey-kuan.github.io/type-D-asymptotics-lean/blueprint/
https://jeffrey-kuan.github.io/type-D-asymptotics-lean/blueprint/dep_graph_document.html
https://jeffrey-kuan.github.io/type-D-asymptotics-lean/docs/

The entire paper was read by myself and I take professional responsibility for the content of this paper.

## 1.6 Acknowledgements

I would like to gratefully thank The Ohio State University's Generative AI initiative, specifically OTDI (Office for Technology and Digital Innovation), ASC Tech (Arts and Sciences Technology); and The Ximera Foundation for providing a Claude Max subscription.

I acknowledge that some of this work was done on devices belonging to Texas A&M University.

Helpful discussions about orthogonal polynomial duality and Boltzmann–Gibbs principle occurred at "Recent Developments in Stochastic Duality" in Eindhoven, with support from EURANDOM and the NWO grant 613.001.753 "Duality for interacting particle systems." Unfortunately, since it occurred in 2022, I no longer remember who I talked to, but any attendee of that conference should consider themselves appreciated.

I would also like to thank Patrik Ferrari and Harvey Weinberger for helpful discussions in Edinburgh, at the workshop "Universality in the Kardar-Parisi-Zhang class and random matrix theory."

Finally, Dr. Taylor Alison Swift deserves acknowledgements for "Female Rage: The Musical (TM)", which helped me remain in academia.

*Conflict of interest disclosure.*

The accessibility work in this paper was done by the author's accessibility company, "Botting Accessibility With LaTeX" (B.A.W.L.) — formerly "Tailor Swift Bot," restructured and renamed at The Ohio State University's request due to obvious copyright issues related to a famous singer–songwriter. B.A.W.L. is a University Technology Commercialization Company (UTCC) as defined in Chapter 3335-13 of The Ohio State University's bylaws and rules. The views and products of B.A.W.L. are those of the company and do not necessarily reflect the views of, nor are they endorsed or approved by, The Ohio State University.

## 2 The model, the measure, and the duality

We first record the three structures on which the entire analysis rests: the generator, the reversible product measure, and the orthogonal $q$–Krawtchouk self–duality. The duality is the decisive one — it is what turns a question about the full interacting field into a question about one or two dual particles, and both regimes are ultimately solved by the same few–body computation.

For a local function $f$ of $\eta \in \{0,1,2,3\}^{\mathbb{Z}}$ the generator is

$$\mathcal{L}f(\eta) = \sum_{x\in\mathbb{Z}} \sum_{(a,b)\to(a',b')} r\left((a,b)\to(a',b')\right) \mathbf{1}\{\omega_x = a, \omega_{x+1} = b\}\left[f(\eta_{a'b'}^{x,x+1}) - f(\eta)\right], \qquad (4)$$

the inner sum over the transitions of (3), $\eta_{a'b'}^{x,x+1}$ denoting $\eta$ with $(\omega_x, \omega_{x+1})$ reset to $(a', b')$. The finite–$n$ rate matrix is (2), displayed in [48, Sec. 2.1] (constructed from the quantum group for $n = 2,3$ in [45]); after the time rescaling by $q^{2n}$ its entrywise $n \to \infty$ limit is (3), which one checks directly to be a Markov generator with the stated conservation laws for every $q \in (0,1)$.

### Proposition 2.1 (reversible measure; [45], Prop. 1.3,[48, Sec. 3.1])

The type D ASEP is reversible with respect to the product blocking measure

$$\nu(\eta) = \mu_{\alpha_1}(\eta_1)\,\mu_{\alpha_2}(\eta_2), \qquad \mu_\alpha(\xi) = \prod_{x\in\mathbb{Z}} \alpha^{\xi_x}\, q^{-2x\xi_x} \quad (\alpha > 0), \qquad (5)$$

a product over both sites and species, carrying the $q^{-2x}$ density tilt, the two species independent.

### Lemma 2.2 (local detailed balance)

For $\nu$ and adjacent sites $(x, x+1)$, the unnormalised two–point weights satisfy

$$\nu(0,3) = q^{-4}\nu(3,0), \qquad \nu(1,2) = \nu(2,1) = q^{-2}\nu(3,0). \qquad (6)$$

*Proof.* By (5) the weight on $\{x, x+1\}$ factorises over the two sites and two species; a species–$i$ particle at site $y$ contributes $\alpha_i q^{-2y}$. Moving the bound pair from $x$ to $x+1$ multiplies the weight by $q^{-2}\cdot q^{-2} = q^{-4}$; moving one species–1 particle from $x$ to $x+1$ and one species–2 particle from $x+1$ to $x$ leaves a single $q^{-2}$ deficit. These are precisely the detailed–balance identities of (3) for $\nu$. ☐

### Remark 2.3 (parameter range and the locked limit)

Although $n$ indexes finite–dimensional representations of $\mathfrak{so}_{2n+2}$, the finite–$n$ generator (2) extends analytically in $n$ and remains a bona fide Markov generator for a range of <u>negative</u>

$n$: the binding rates $(1,2) \to (3,0)$, $(3,0) \to (1,2)$ are the constraints, both proportional to $2 - q^{2n-2}(1-q^2)$ and hence vanishing at $n_{\text{crit}} = \log(2q^2/(1-q^2))/(2\log q)$, so the process is stochastic for $n > n_{\text{crit}}$, with $n_{\text{crit}} < 0$ exactly when $q > 1/\sqrt{3}$ (i.e. $2q^2/(1-q^2) > 1$). At $q = 1$ the split/merge rates vanish, bound pairs never break, $N_1 \equiv N_2$, and the species are perfectly locked.

## 2.1 The orthogonal $q$–Krawtchouk self–duality

Each site holds at most one particle of each species, so the single–site, single–species marginal is the two–point specialisation of the $q$–Krawtchouk family. With local fugacity $t_x := \alpha q^{-2x}$ and density $\rho_x = t_x/(1+t_x)$, the degree–$\leq 1$ orthogonal polynomials of $\mu_\alpha$ are

$$\kappa_0(\eta_x) \equiv 1, \qquad \kappa_1(\eta_x) = \frac{(1+t_x)\eta_x - t_x}{\sqrt{t_x}} = \frac{\eta_x - \rho_x}{\sqrt{\rho_x(1-\rho_x)}}. \qquad (7)$$

*Definition 2.4 (orthonormal local basis)*

For a dual configuration $\xi$ specifying which (site, species) pairs carry a dual particle, set

$$D^{\text{loc}}(\xi,\eta) = \prod_{(x,i):\, \xi_{i,x}=1} \kappa_1(\eta_{i,x}), \qquad (8)$$

a polynomial in $\eta$ of degree $|\xi|$ (the dual mass). Throughout, $L^2(\nu)$ carries the inner product $\langle f, g\rangle_\nu := \mathbb{E}_\nu[fg] = \int fg \; \mathrm{d}\nu$. By the product structure of $\nu$ and the single–site orthonormality of (7), the family $\{D^{\text{loc}}(\xi,\cdot)\}$ is orthonormal in $L^2(\nu)$. It is the natural basis for the fluctuation fields: mass–one elements are the centred density fields, and the different–species mass–two element at one site is the bound–pair mode.

*Theorem 2.5 (orthogonal q–Krawtchouk self–duality)*

For configurations $\eta, \xi$ and $i \in \{1,2\}$ let $N^-_{x-1}(\xi_i)$ be the number of species–$i$ dual particles strictly to the left of $x$ and $N^+_{x+1}(\eta_i)$ the number of species–$i$ particles of $\eta$ strictly to the right of $x$, and set

$$D_{\alpha_1,\alpha_2}(\xi,\eta) = \prod_{i=1,2} \prod_{x:\, \xi_{i,x}=\eta_{i,x}=1} \left(1 - \frac{q^{2\left(x - N^-_{x-1}(\xi_i) + N^+_{x+1}(\eta_i)\right)}}{\alpha_i}\right), \qquad (9)$$

the two–point specialisation of the $q$–Krawtchouk product of [48], Thm. 3.1.

1. ([48], Thm. 3.1) For parameters $(q,n,\delta) = (q,2,0)$ and $(q,3,0)$, the type D ASEP is self–dual with respect to (9).

2. At $n = \infty$, i.e. for the rates (3): the two–site interlacing $\mathcal{L}D = D\mathcal{L}^{\mathsf{T}}$ holds identically in $(q,\alpha_1,\alpha_2)$ — a finite rational identity, which we have verified by computer algebra — and the induction in the proof of [48], Thm. 3.1, eqs. (16)–(19), which uses only the two–site case and the fugacity–shift covariance of (9), extends it to any number of sites. This is the $n = \infty$ case of the conjecture of [48], Rmk. 4; we claim no more than the verification, the method being that of [48]. In the lineage of orthogonal stochastic dualities from $U_q$–representations [61], [62], this is the type D instance.

The triangular self–duality of the type D ASEP is prior work: it is the original duality of [45], the type D analogue of the Schütz and Borodin–Corwin–Sasamoto $q$–exponential dualities [49], [57]. We record only the form we use at $n = \infty$.

*Corollary 2.6 (triangular form at $n = \infty$)*

At $n = \infty$ the type D ASEP is self–dual with respect to the triangular function

$$D^{\text{tri}}(\xi,\eta) \;=\; \mathbf{1}\{\xi \subseteq \eta\} \prod_{i=1,2} \; \prod_{x:\,\xi_{i,x}=1} q^{2\left(x - N^{-}_{x-1}(\xi_i) + N^{+}_{x+1}(\eta_i)\right)}. \qquad (10)$$

On configurations $\eta$ with finitely many particles to the right — in particular on the step initial condition and its evolution — the factors of (10) are uniformly bounded for each fixed dual mass, and (10) is the form of the duality used in §8.

*Proof.* Self–duality with respect to (9) holds identically in $(\alpha_1, \alpha_2)$, hence coefficient–wise in $\alpha_1^{-1}, \alpha_2^{-1}$; and since the dual generator preserves the particle numbers $(|\xi_1|, |\xi_2|)$, the coefficients of $\alpha_1^{-|\xi_1|}\alpha_2^{-|\xi_2|}$, glued across the dual sectors, again form a duality matrix. Extracting the leading coefficients as $\alpha_i \to 0$ yields self–duality with respect to (10). (Equivalently, the interlacing $\mathcal{L}D^{\text{tri}} = D^{\text{tri}}\mathcal{L}^{\mathsf{T}}$ is a direct check, which we have also verified by computer algebra.) □

The following is [55], Lem. 2.1, which we restate for an arbitrary measure $\varpi$ orthogonalising the duality family — ACR state it for the standard reversible measure $\nu$, whereas in §6 we apply it with the sector–reweighted $\varpi \neq \nu$ of Proposition 2.8 below. The proof, identical to theirs and using only orthogonality and the duality relation (not reversibility of $\varpi$), is a single line, included for completeness.

*Lemma 2.7 (duality covariance; after [55], Lem. 2.1)*

Let $\{D(\xi,\cdot)\}$ be a family of duality functions for the process which is orthogonal in $L^2(\varpi)$ for some measure $\varpi$, $\int D(\xi,\eta)D(\xi',\eta)\,\mathrm{d}\varpi = \delta_{\xi,\xi'}a(\xi)$. Then for all $\xi, \xi'$ and $t \geq 0$,

$$\int \mathbb{E}_\eta[D(\xi,\eta_t)]\, D(\xi',\eta)\,\mathrm{d}\varpi(\eta) = p_t(\xi,\xi')\,a(\xi').$$

*Proof.* Insert the duality identity $\mathbb{E}_\eta[D(\xi,\eta_t)] = \sum_{\xi''} p_t\,(\xi,\xi'')D(\xi'',\eta)$, multiply by $D(\xi',\eta)$, and integrate against $\varpi$. By orthogonality $\int D(\xi'',\eta)D(\xi',\eta)\,\mathrm{d}\varpi = \delta_{\xi'',\xi'}a(\xi')$, so only $\xi'' = \xi'$ survives. The interchange of sum and integral is justified by the finite mass of $\xi$ (only finitely many $\xi''$ are reachable in the particle–conserving dual). □

We record the orthogonality of the duality functions in the form used in §6. On a finite lattice $\Lambda$ define the sector–reweighted blocking measure and norm functions ($N = N(\zeta)$ the particle number, up to normalisation)

$$\varpi_\alpha(\zeta) \;\propto\; \mu_\alpha(\zeta)\, q^{N(N-1)}\, (\alpha q^{2N-2|\Lambda|};q^2)_\infty, \quad g_\alpha(\zeta) \propto\; q^{2N(N-1)} (\alpha q^{2N-2|\Lambda|};q^2)_\infty (\alpha q^{-2N};q^2)_\infty, \qquad (11)$$

and let $\bar{D}$ denote (9) read with the <u>process</u> configuration in the $N^-$ slot and the dual in the $N^+$ slot,

$$\bar{D}(\xi,\eta) = \prod_{i=1,2} \; \prod_{x:\,\xi_{i,x}=\eta_{i,x}=1} \left(1 - \frac{q^{2\left(x - N^-_{x-1}(\eta_i) + N^+_{x+1}(\xi_i)\right)}}{\alpha_i}\right).$$

*Proposition 2.8 (orthogonality; [63], Thm. 3.2, [48, Sec. 3.1])*

For $\alpha_1, \alpha_2 \in (0, q^{2|\Lambda|})$, set $\varpi := \varpi_{\alpha_1} \otimes \varpi_{\alpha_2}$. Then $\varpi$ is a reversible measure, and the functions $\bar{D}$ are orthogonal in $L^2(\varpi)$:

$$\sum_\eta \varpi_{\alpha_1}(\eta_1)\, \varpi_{\alpha_2}(\eta_2)\, \bar{D}(\xi,\eta)\, \bar{D}(\xi',\eta) \;=\; \delta_{\xi,\xi'}\, a(\xi), \qquad a(\xi) = \prod_{i=1,2} \frac{g_{\alpha_i}(\xi_i)}{\varpi_{\alpha_i}(\xi_i)}. \qquad (12)$$

This is the two–species product of the single–species $q$–Krawtchouk orthogonality [63], Thm. 3.2, as noted in [48, Sec. 3.1]; being a statement about the duality functions and the measure, it is independent of $n$. Consequently Lemma 2.7 applies with $(\varpi, \bar{D}, a)$.

*Remark 2.9 (orientation, density window, infinite volume)*

Three points on the use of Proposition 2.8 in regime (A). <u>Orientation:</u> $\bar{D}$ places the process in the <u>left</u> counts $N^-(\eta_i)$, which are a.s. finite under $\nu$ on $\mathbb{Z}$; the opposite orientation (9) degenerates there ($q^{2N^+(\eta)} = 0$ a.s.), and the fix is to transpose the slots, not the lattice (the reflected generator fails the interlacing). <u>Fugacities:</u> in §6 the reweighted measure $\varpi$ is instantiated at the <u>compensated</u> fugacities $\beta_i = \alpha_i/(1+\alpha_i) = \rho_i$ of Lemma 6.9; since $\beta_i < 1$ for every density $\rho_i \in (0,1)$, the window hypothesis $\beta_i < q^{2|\Lambda|} \to 1$ of Proposition 2.8 holds for all large $N$ at <u>every</u> pair of densities — above, below, or straddling $1/2$. <u>Infinite volume:</u> the replacement of $\Lambda = [-KN, KN]$ by $\mathbb{Z}$ in the correlation functions, and the uniformity of the resulting constants, is deferred to future work. In regime (B) we use the triangular duality of Corollary 2.6 instead, where none of this enters.

Lemma 2.7 reduces every space–time correlation of degree–$k$ fields to a single $k$–particle dual kernel evaluation — the orthogonal–duality reduction of few–body correlations of [55], [56], and the mechanism that makes the fluctuation theory a few–body problem. In this paper it is applied at $k = 2$, to the inter–species cross–noise (§6).

## 3 Exact current decoupling and vanishing transport coefficients

The structural fact behind every decoupling result in this paper is elementary and exact, holding at the level of the microscopic rates and at <u>every</u> $n$: each species' current does not see the other. We prove this, with its consequences — autonomous single–species marginals, a diagonal hydrodynamic flux Hessian, and a coupling confined to correlated jumps — in Proposition 3.1, and then that the two equilibrium cross–transport coefficients vanish identically (Proposition 3.2). Everything afterwards is the question of what residual coupling can still survive once these channels are closed.

## Proposition 3.1 (exact current decoupling)

For every $n \in \mathbb{N} \cup \{\infty\}$ the microscopic species–$i$ current across a bond $(x, x+1)$ equals the single–species ASEP current in $\eta_i$ alone,

$$j_i = r_R\, \eta_{i,x}(1-\eta_{i,x+1}) - r_L\,(1-\eta_{i,x})\eta_{i,x+1}, \qquad r_R = q^{-1}\beta_n,\ \ r_L = q\,\beta_n,\ \ \beta_n = q^{1-2n} + q^{2n-1}, \qquad (13)$$

with no dependence on the other species' occupancy; at $n = \infty$, after rescaling time by $q^{2n}$, the rates become $r_R = 1$, $r_L = q^2$ (those of (3)). Consequently, for every $n$:

1. each marginal density field evolves as an autonomous single–species ASEP;
2. the macroscopic flux $\mathbb{E}_{\nu_\rho}[j_i] = (r_R - r_L)\,\rho_i(1-\rho_i)$ is a function of $\rho_i$ alone, so all cross derivatives vanish, $\partial_{\rho_1} j_2 = \partial_{\rho_1}^2 j_2 = \partial_{\rho_1}\partial_{\rho_2} j_2 = 0$, and the two–species hydrodynamic flux Hessian is diagonal;
3. the inter–species coupling is carried entirely by correlated jumps — bound–pair translations, swaps, and binding/splitting — none of which alters (13).

*Proof*. We compute the signed rate of species–1 transfer across $(x, x+1)$, $j_1 = \sum_{\text{transitions}} r(\cdot)\,\Delta\eta_1$, over each species–2 occupancy of the bond, and show it is independent of that occupancy; species 2 is symmetric.

The limit $n = \infty$ (rates (3)). Rightward: $(1{,}0) \to (0{,}1)$, the pair hop $(3{,}0) \to (0{,}3)$, and $(3{,}2) \to (2{,}3)$ (species 1 sliding past the pair) each at rate 1, and from $(1{,}2)$ the merge $(1{,}2) \to (0{,}3)$ at rate $\varepsilon$ plus the swap $(1{,}2) \to (2{,}1)$ at rate $q^2$, with $\varepsilon + q^2 = 1$; all four occupancies give $r_R = 1$. Leftward: $(0{,}1) \to (1{,}0)$, $(2{,}3) \to (3{,}2)$, $(2{,}1) \to (1{,}2)$ at rate $q^2$, and from $(0{,}3)$ the pair hop $(0{,}3) \to (3{,}0)$ at rate $q^4$ plus the split $(0{,}3) \to (1{,}2)$ at rate $q^2\varepsilon$, with $q^4 + q^2\varepsilon = q^2$; all give $r_L = q^2$. (The channels $(2{,}1) \to (0{,}3)$, $(0{,}3) \to (2{,}1)$ and the mixed hops $(3{,}1) \leftrightarrow (1{,}3)$ leave $\eta_1$ on the bond unchanged.)

Finite $n$ (the rates (2), i.e. the generator of [48, Sec. 2.1], constructed in [45] for $n = 2{,}3$). Four further channels are present that vanish in the rescaled limit: the left merges $(1{,}2), (2{,}1) \to (3{,}0)$ of rate $2 - q^{2n-2}(1-q^2)$ and the left splits $(3{,}0) \to (1{,}2), (2{,}1)$ of rate $q^{-2}(2 - q^{2n-2}(1-q^2))$. With $\sigma_n = (q^{n-1} - q^{1-n})^2 = q^{2n-2} - 2 + q^{2-2n}$ as in (1), the species–1 rightward rate, case by case in the species–2 occupancy, reads

$$
\begin{aligned}
&(1,0):\ q^{-1}\beta_n = q^{-2n} + q^{2n-2};\\
&(3,0):\ \underbrace{q^{-2}\sigma_n}_{(3,0)\to(0,3)} + \underbrace{q^{2n-2} - q^{2n-4} + 2q^{-2}}_{(3,0)\to(2,1)} = q^{-2n} + q^{2n-2};\\
&(1,2):\ \underbrace{2 + q^{-2n}(1-q^2)}_{(1,2)\to(0,3)} + \underbrace{\sigma_n}_{(1,2)\to(2,1)} = q^{-2n} + q^{2n-2};\\
&(3,2):\ q^{-1}\beta_n,
\end{aligned}
$$

and the leftward rates all equal $q\beta_n = q^{2-2n} + q^{2n}$ (in the $(0,3)$ background, the pair hop $q^2\sigma_n$ plus the split $2q^2 + q^{2-2n} - q^{4-2n}$; in the $(2,1)$ background, the swap $\sigma_n$ plus the left merge $2 - q^{2n-2}(1-q^2)$). Hence $r_R = q^{-1}\beta_n, r_L = q\beta_n$ for every $n$, giving (13); the entrywise $q^{2n}$–rescaled $n \to \infty$ limit recovers (3). We verified the full statement and this limit by computer algebra on the $16 \times 16$ generator of [48, Sec. 2.1].

Consequences. (a) is immediate from (13). For (b), the hydrodynamic flux $\mathbb{E}_{\nu_\rho}[j_i] = (r_R - r_L)\rho_i(1-\rho_i)$ is a function of $\rho_i$ alone, so every $\rho_j$–derivative with $j \neq i$ vanishes, as does $\partial_{\rho_1}\partial_{\rho_2} j_i$. (c) is the residual: the only transitions correlating the two species — bound–pair hops, swaps, and the merge/split channels (including the extra ones present at finite $n$) — are precisely those absent from (13), and by the tally above none alters the marginal rate. ☐

For the transport coefficients introduce the bond function

$$
V_x := [\mathbf{1}_{(3,0)} + q^4\mathbf{1}_{(0,3)} - q^2\mathbf{1}_{(1,2)} - q^2\mathbf{1}_{(2,1)}]_{x,x+1}, \qquad \mathbf{1}_{(a,b)} = \mathbf{1}\{\omega_x = a, \omega_{x+1} = b\}, \qquad (14)
$$

the carré du champ of the only events moving both species across the bond (bound–pair hops, sign $+$; swaps, sign $-$).

## Proposition 3.2 (vanishing cross coefficients)

$\mathbb{E}_\nu[V_x] = 0$ (equivalently, the equilibrium cross–mobility $\sigma_{12} = 0$), and the static cross–compressibility $C_{12} := \sum_r \langle \delta\rho_1(0)\delta\rho_2(r)\rangle_\nu = \mathrm{Cov}(N_1, N_2)/L$ vanishes identically.

*Proof.* By Lemma 2.2, $\mathbb{E}_\nu[V_x] = \nu(3,0) + q^4\nu(0,3) - q^2\nu(1,2) - q^2\nu(2,1) = \nu(3,0)[1 + 1 - 1 - 1] = 0$. The compressibility statement is the product structure of $\nu$: the two species are independent under $\nu$, so all equal–time cross–correlations vanish and $C_{12} = 0$. ☐

## Remark 3.3 (interpretation)

By Proposition 3.1 the asymmetry — and hence the KPZ nonlinearity — lives only in the marginals, which are autonomous ASEPs, while the inter–species coupling is purely a noise effect; by Proposition 3.2 that coupling is invisible to both equilibrium transport coefficients. Physically, a bound pair carries one particle of each species together, but it hops with exactly the bias the two single–species currents already have, so it injects correlated fluctuations without shifting any mean flow: coupling in the noise, not in the drift. These two facts underlie the decoupling in both regimes. The cancellation $1 + 1 - 1 - 1 = 0$ of Proposition 3.2 recurs three times in §6.

# 4 Local central limit theorems for the dual coordinates

The dual processes below reduce, after passing to a relative coordinate, to one–dimensional walks. We collect the local CLTs they require. Throughout, a driftless finite–range walk on $\mathbb{Z}$ means a continuous–time, irreducible, aperiodic, mean–zero walk with bounded jump range.

## Lemma 4.1 (free and perturbed kernel bounds)

Let $R^0$ be the homogeneous symmetric nearest–neighbour walk of total rate $a > 0$. Then $\mathbb{P}(R^0(t) = r) \le C/\sqrt{1+t}$ for all $t \ge 0, r \in \mathbb{Z}$, with $C = C(a)$ — the classical local central limit / on–diagonal heat–kernel estimate for the continuous–time simple random walk (the special case $m \equiv 1$ of (15) below; see also [64]). Moreover, let $\tilde{R}$ be an irreducible walk on $\mathbb{Z}$ (or on a state space graph–isomorphic to $\mathbb{Z}$, such as $\mathbb{Z}\backslash\{0\}$ relabelled) which is reversible with respect to a measure $m$ with $c_1 \le m \le c_2$, has jump range $\le \varrho$ and total exit rates $\le \Lambda$, and whose nearest–neighbour conductances are bounded below, $m(x)\, r(x, x+1) \ge \delta > 0$ for all $x$. Then

$$\sup_r \mathbb{P}\big(\tilde{R}(t) = r\big) \le \frac{C}{\sqrt{1+t}}, \qquad C = C(c_1, c_2, \delta, \Lambda, \varrho). \qquad (15)$$

Every perturbed walk in this paper satisfies the hypotheses: the no–merge relative walk on $\mathbb{Z}\backslash\{0\}$ of Lemma 4.2 below and of Lemma 7.7 in §7 is reversible with respect to counting measure (the swap channel is symmetric), with $\delta \ge q^2$, so its constants are uniform as $q = q_N \to 1$; the full relative walk of §8 is reversible with respect to $m(0) = q^{-2}, m \equiv 1$ otherwise, with $\delta = \min(\varepsilon, q^2) > 0$ at fixed $q$.

*Proof*. The bound for $R^0$ is the special case $m \equiv 1$ of (15), which we now prove. The bound is trivial for $t \le 1$, so let $t \ge 1$. On $\ell^2(m)$ the generator is self–adjoint with Dirichlet form $\mathcal{E}(f) = 1/2 \sum_{x,y} m(x) r(x,y)(f(y) - f(x))^2 \;\ge\; \delta \sum_x (f(x+1) - f(x))^2$. On $\mathbb{Z}$ the discrete Nash inequality $\| f \|_2^6 \le 4 \;\| \nabla f \|_2^2 \| f \|_1^4$ holds: from $f(x)^2 = \sum_{y<x} \nabla (f^2)(y)$ and Cauchy–Schwarz, $\| f \|_\infty^2 \le 2 \;\| \nabla f \|_2 \| f \|_2$, so $\| f \|_2^2 \le \| f \|_1 \| f \|_\infty \le \| f \|_1 \;(2 \;\| \nabla f \|_2 \| f \|_2)^{1/2}$, and rearranging gives the claim. With the measure bounds $c_1 \le m \le c_2$ this transfers to $\| f \|_{2,m}^6 \le C\, \mathcal{E}(f) \;\; \| f \|_{1,m}^4$, a Nash inequality with dimension parameter $d = 1$ for the Dirichlet form of $\tilde{R}$. By the equivalence of Nash inequalities and on–diagonal kernel bounds for symmetric Markov semigroups [65], $\sup_{x,y} p_t(x,y)/m(y) \le C t^{-1/2}$, and $m \le c_2$ gives (15); all constants depend only on $(c_1, c_2, \delta, \Lambda, \varrho)$. □

The Edwards–Wilkinson kernel requires, in addition, control of the slow sojourn at the origin created by the small split rate. Recall $\varepsilon = 1 - q^2$ and set $\nu_{\rm sp} := 2q^2\varepsilon$. The relevant relative walk is $R := X_2 - X_1$, the difference of the positions $X_1, X_2$ of the two different–species dual particles (the pair analysed in §5). Off $\{-1,0,1\}$ it agrees with a rate–$(1 + q^2)$ symmetric walk; its origin $R = 0$ (the bound pair) is sticky, held for an $\mathrm{Exp}(\nu_{\mathrm{sp}})$ time and re–entered only from $\pm 1$ at rate $\varepsilon$.

## Lemma 4.2 (defected marginal local CLT)

Let $R$ be the walk just described, started at $R(0) = 0$. For $\nu_{\mathrm{sp}} t \le K$,

$$\mathbb{P}(R(t) = r) \;\le\; \frac{C}{\sqrt{1+t}} \;+\; \delta_{r,0}\, e^{-\nu_{\mathrm{sp}} t}, \qquad C = C(K,q). \qquad (16)$$

*Proof.* Write $\nu = \nu_{\mathrm{sp}}$, so $\varepsilon = \nu/(2q^2)$, and recall $\nu t \le K$, $\nu \le 2$; all constants are $C = C(K,q)$. For $t \le 1$ the bound is trivial, so let $t \ge 1$. We argue by cases on whether $r = 0$: for $r \ne 0$ the claim (16) reads $\mathbb{P}(R(t) = r) \le C/\sqrt{1+t}$, established below as (18), and for $r = 0$ it reads $\mathbb{P}(R(t) = 0) \le e^{-\nu t} + C/\sqrt{1+t}$, established as (19). We bound $R$ by splitting its path into excursions away from the sticky origin. The expected number of these excursions is at most $K$ — this is where the window hypothesis $\nu_{\mathrm{sp}} t \le K$ enters, as quantified in the Occupation of the adjacent set step below — while a single excursion is controlled by comparing $R$ with two auxiliary walks. The no–merge walk $\hat{R}$ is $R$ with the merge channel deleted: on $\mathbb{Z}\backslash\{0\}$ it agrees with $R$, but once it leaves the origin it never returns, so it is a perturbed walk of the type of Lemma 4.1 and satisfies (15) with constants uniform in $q \to 1$. Killing $\hat{R}$ at rate $\varepsilon$ while in $\{\pm 1\}$ (the merge rate) gives a sub–probability walk whose kernel is dominated by that of $\hat{R}$, since killing only removes mass. A single no–merge excursion of $R$ agrees in law with this killed walk, so for $r \ne 0$ and any starting point in $\{\pm 1\}$,

$$\mathbb{P}_{\pm 1}(R(u) = r,\ \text{no merge in } [0,u]) = \mathbb{E}_{\pm 1}\big[\mathbf{1}\{\hat{R}(u) = r\}\, e^{-\varepsilon \Lambda_u}\big] \;\le\; \frac{C}{\sqrt{1+u}}, \qquad (17)$$

$\Lambda_u$ the occupation of $\{\pm 1\}$ by $\hat{R}$.

Atom. On $\{t < \tau_1\}$, $\tau_1 \sim \mathrm{Exp}(\nu)$ the first split, the walk has never left 0: this is the term $\delta_{r,0} e^{-\nu t}$.

Occupation of the adjacent set. Let $g(t) := \int_0^t \mathbb{P}\,(R(u) \in \{\pm 1\})\, \mathrm{d}u$. Exits from 0 occur at rate $\nu$ while at 0 (the two splits; pair hops do not change $R$), so the expected number of exits from 0 — equivalently, of excursions — by time $t$ is $\nu \int_0^t \mathbb{P}\,(R(u) = 0)\, \mathrm{d}u \le \nu t \le K$, the window hypothesis. By the strong Markov property at each exit and (17), each excursion contributes to $g$ at most $\int_0^t C\,(1+v)^{-1/2}\, \mathrm{d}v \le C\sqrt{t}$ in expectation, whence $g(t) \le (1+K)\, C\sqrt{t}$.

Off the origin. Fix $r \ne 0$. On $\{R(t) = r\}$ the time $t$ lies in an excursion, begun at an exit from 0; the exit times form a point process of intensity $\nu\, \mathbf{1}\{R(u) = 0\}\, \mathrm{d}u$, and given an exit at $u$ the probability that its excursion still runs at time $t$ with $R(t) = r$ is bounded by (17). Hence

$$\mathbb{P}(R(t) = r) \;\le\; \nu \int_0^t \mathbb{P}\,(R(u) = 0)\, \frac{C}{\sqrt{1+t-u}}\; \mathrm{d}u,$$

and we insert $\mathbb{P}(R(u)=0) \le e^{-\nu u} + \varepsilon\, g(u) \le e^{-\nu u} + C\varepsilon\sqrt{u}$ (the two events being "never left" and "last merge before $u$", the latter of intensity $\varepsilon\mathbf{1}\{R \in \{\pm 1\}\}$). The exponential part: on $u \le t/2$ it is at most $C(1+t)^{-1/2}\int_0^\infty \nu\, e^{-\nu u}\,\mathrm{d}u = C(1+t)^{-1/2}$; on $u > t/2$ it is at most $\nu e^{-\nu t/2} \cdot 2\sqrt{t} \le C(1+t)^{-1/2}$, since $\nu\sqrt{t}\,\sqrt{1+t} \le 2\nu(1+t) \le 2(\nu+K)$. The occupation part is at most $C\nu\varepsilon\int_0^t \sqrt{u}\,(1+t-u)^{-1/2}\,\mathrm{d}u \le C\nu\varepsilon\sqrt{t}\cdot 2\sqrt{t} = C\nu\varepsilon\, t$, and multiplying by $\sqrt{1+t}$, $\nu\varepsilon\, t\sqrt{1+t} \le 2\nu\varepsilon t^{3/2} = 2(\nu t)\cdot\varepsilon\sqrt{t} = 2(\nu t)\cdot\frac{\sqrt{\nu}\,\sqrt{\nu t}}{2q^2} \le \frac{\sqrt{2}\,K^{3/2}}{q^2} \le C$, so this part too is $\le C(1+t)^{-1/2}$. Adding the exponential and occupation parts, we obtain, for every $r \ne 0$,

$$\mathbb{P}(R(t)=r) \;\le\; \frac{C}{\sqrt{1+t}}. \qquad (18)$$

<u>At the origin.</u> Decomposing $\{R(t)=0\}$ on the last merge before $t$ (or none),

$$\begin{aligned}\mathbb{P}(R(t)=0) &= e^{-\nu t} + \varepsilon\int_0^t \mathbb{P}\left(R(u)\in\{\pm 1\}\right) e^{-\nu(t-u)}\,\mathrm{d}u \;\le\; e^{-\nu t} + \varepsilon\int_0^t \frac{C}{\sqrt{1+u}}\,\mathrm{d}u \\ &\le\; e^{-\nu t} + C\varepsilon\sqrt{t},\end{aligned}$$

using the off–origin bound just proved at $r=\pm 1$; and $\varepsilon\sqrt{t}\,\sqrt{1+t} \le 2\varepsilon t = \nu t/q^2 \le K/q^2$ shows $C\varepsilon\sqrt{t} \le C(1+t)^{-1/2}$, i.e.

$$\mathbb{P}(R(t)=0) \;\le\; e^{-\nu t} + \frac{C}{\sqrt{1+t}}. \qquad (19)$$

<u>Conclusion.</u> The bound (16) now follows from the two cases. If $r \ne 0$ then $\delta_{r,0}=0$ and (18) is exactly (16). If $r=0$ then $\delta_{r,0}=1$ and (19) is exactly (16), its atom $e^{-\nu t}$ being the never–left contribution of the <u>Atom</u> step. This proves (16) for all $r \in \mathbb{Z}$ at $t \ge 1$, and the case $t \le 1$ is the trivial bound noted at the start. ☐

The relative coordinate is now controlled; we turn to $S = X_1 + X_2$. The Fourier argument used for $R^0$ is unavailable, because $S$ and $R$ are driven by the same jumps, so once the $R$–path is fixed the increments of $S$ are neither independent nor identically distributed. We use anti–concentration instead. The mechanism: the $R$–path, even together with all jump times, does not determine $S$. While the two particles are apart, a relative step can be realised either by particle 1 moving or by particle 2 moving, the two realisations changing $S$ in opposite directions; and a bound–pair hop (which leaves $R$ fixed) changes $S$ by $\pm 2$. Each such choice is an independent coin whose two outcomes have probability bounded away from 0 and 1. Conditioning on all the data <u>except</u> these choices, $S(t)$ is a deterministic value plus a sum of $M$ independent $\pm$ steps; the Kolmogorov–Rogozin inequality (Lemma 4.3), an anti–concentration bound on the largest atom of a sum of independent variables, spreads such a sum over a window of width $\sqrt{M}$, and jumps occur at rate $\ge 1$, so $M \gtrsim t$. We make this precise; the data conditioned on is named the <u>unsigned skeleton</u> in the lemma.

### Lemma 4.3 (Kolmogorov–Rogozin anti–concentration; [66, Ch. III])

For an integer–valued random variable $Y$, let $Q(Y) := \sup_{x\in\mathbb{Z}} \mathbb{P}(Y = x)$ denote its largest atom. There is a universal constant $C$ such that, for any independent integer–valued $Y_1, \dots, Y_n$,

$$Q(Y_1 + \cdots + Y_n) \;\le\; C\Big(\sum_{j=1}^{n} \big(1 - Q(Y_j)\big)\Big)^{-1/2}.$$

In particular, if each $Y_j$ takes two values, each of probability at least $\delta$, then $Q(Y_1 + \cdots + Y_n) \le C(\delta n)^{-1/2}$.

### Lemma 4.4 (conditional concentration for the sum coordinate)

Let $S = X_1 + X_2$ be the sum coordinate, and write the jumps of the pair $(X_1, X_2)$ at times $t_1 < t_2 < \cdots$, with increments $\Delta R_k := R(t_k) - R(t_k^-)$ and $\Delta S_k := S(t_k) - S(t_k^-)$. The <u>unsigned skeleton</u> is

$$\mathfrak{S} \;:=\; \big((R(u))_{0\le u\le t}\,,\ \{\,t_k : \Delta R_k = 0\,\}\big), \qquad (20)$$

the path of $R$ together with the times of the pair hops (the $\Delta R = 0$ jumps, which the $R$–path does not reveal). A jump $t_k$ is <u>ambiguous</u> if $\Delta S_k$ is not determined by $\mathfrak{S}$ — equivalently, two values of $\Delta S_k$ are consistent with $\mathfrak{S}$ — and $M = M(\mathfrak{S})$ is the number of ambiguous jumps in $[0, t]$. Then for every $s' \in \mathbb{Z}$,

$$\mathbb{P}(S(t) = s' \mid \mathfrak{S}) \;\le\; \frac{C}{\sqrt{1 + M}},$$

and $M$ stochastically dominates a Poisson variable of mean $t$, so $\mathbb{P}(M < t/2) \le e^{-ct}$.

*Proof.* Given $\mathfrak{S}$, $S(t) = m(\mathfrak{S}) + \sum_{j=1}^{M} \eta_j$ with $m(\mathfrak{S})$ determined by the skeleton and the $\eta_j$ — the values $\Delta S_k$ at the ambiguous jumps — conditionally independent: on $\{|R| \ge 2\}$, and on the outward steps from $\{\pm 1\}$, a step $\Delta R = +1$ is either "species 1 left" ($\Delta S = -1$, conditional probability $q^2/(1 + q^2)$) or "species 2 right" ($\Delta S = +1$), and symmetrically for $\Delta R = -1$; a pair hop has $\Delta S = \pm 2$ with conditional probabilities $(1, q^4)/(1 + q^4)$. (Merges, splits and swaps determine $\Delta S$ and are unambiguous.) Each $\eta_j$ is non–degenerate with both values of probability in $[\delta, 1 - \delta]$, $\delta = \delta(q) > 0$, so by Lemma 4.3 applied to the $M$ ambiguous increments, $\mathbb{P}(S(t) = s' \mid \mathfrak{S}) \le C(\sum_{j=1}^{M} (1 - Q(\eta_j)))^{-1/2} \wedge 1 \le C(1 + \delta M)^{-1/2}$. For the domination: in every state the total rate of ambiguous events is $\ge 1$ — rate $2(1 + q^2)$ on $\{|R| \ge 2\}$, $1 + q^2$ (the outward steps) at $\{\pm 1\}$, and $1 + q^4$ (the pair hops) at the origin — so, conditionally on any past, ambiguous events fire at rate $\ge 1$, $M$ dominates a rate–1 Poisson count, and $\mathbb{P}(M < t/2) \le e^{-ct}$. □

The fixed–$q$ analysis of §8 requires, beyond the upper bound (15), the exact leading occupation asymptotics of the relative walk, including the invariant–measure weighting at the defect. The following elementary lemma supplies them; its entire content is two exact identities (detailed balance and first–passage decomposition) plus the free first–passage transform and the following classical Tauberian theorem, which converts the small–$\lambda$ behaviour of a Laplace transform into the large–$s$ behaviour of the integral of its density.

### Theorem 4.5 (Karamata's Tauberian theorem; [67], Thm. 1.7.1′, [68, Sec. XIII.5])

Let $p: [0,\infty) \to [0,\infty)$ be locally integrable with Laplace transform $\omega(\lambda)$ $:= \int_0^\infty e^{-\lambda t}\, p(t)\,\mathrm{d}t < \infty$ for all $\lambda > 0$, let $\rho \geq 0$, and let $L$ be slowly varying at infinity. Then

$$\omega(\lambda) \;\sim\; \lambda^{-\rho}\, L(1/\lambda) \quad (\lambda \downarrow 0) \qquad \Leftrightarrow \qquad \int_0^s p\,(t)\,\mathrm{d}t \;\sim\; \frac{s^\rho\, L(s)}{\Gamma(\rho+1)} \quad (s \to \infty).$$

### Lemma 4.6 (occupation–time asymptotics)

Let $X$ be an irreducible, recurrent walk on $\mathbb{Z}$, reversible with respect to a measure $m$ with $c_1 \leq m \leq c_2$ and $m \equiv 1$ off a finite set, and agreeing off a finite set with the symmetric nearest–neighbour walk of rate $a$ per direction. Then for every fixed $r$,

$$\tau_r(s) := \int_0^s \mathbb{P}_0\,(X(t) = r)\,\mathrm{d}t \;\sim\; m(r)\sqrt{\frac{s}{\pi a}} \qquad (s \to \infty).$$

*Proof.* Let $G_\lambda(x,y) = \int_0^\infty e^{-\lambda t}\,\mathbb{P}_x(X(t) = y)\,\mathrm{d}t$, $\lambda > 0$. Two exact identities: detailed balance gives $m(x)G_\lambda(x,y) = m(y)G_\lambda(y,x)$, and the first–passage decomposition at $y$ gives $G_\lambda(x,y) = \mathbb{E}_x[e^{-\lambda T_y}]\,G_\lambda(y,y)$. Combining,

$$G_\lambda(0,r) = \frac{m(r)}{m(0)}\,\mathbb{E}_r\big[e^{-\lambda T_0}\big]\,G_\lambda(0{,}0), \qquad (21)$$

and by recurrence $\mathbb{E}_r[e^{-\lambda T_0}] \to 1$ as $\lambda \downarrow 0$ for each fixed $r$. To anchor $G_\lambda(0{,}0)$, sum (21) over $r$ and use $\sum_r G_\lambda\,(0,r) = \lambda^{-1}$:

$$\frac{1}{\lambda} = \frac{G_\lambda(0{,}0)}{m(0)}\sum_r m\,(r)\,\mathbb{E}_r\big[e^{-\lambda T_0}\big].$$

Let $I$ be a finite interval containing the defect and the origin. From $r$ outside $I$ the walk is free until it first hits $I$ (necessarily at the near edge), and from $\partial I$ it reaches $0$ in a time which is almost surely finite, so $\mathbb{E}_r[e^{-\lambda T_0}] = \varrho(\lambda)^{\mathrm{dist}(r,I)}\,\mathbb{E}_{\partial I}[e^{-\lambda T_0}]$ with $\mathbb{E}_{\partial I}[e^{-\lambda T_0}] \to 1$, where $\varrho(\lambda)$ is the free nearest–neighbour first–passage root, $\lambda + 2a = a(\varrho + \varrho^{-1})$, i.e. $\varrho(\lambda) = 1 - \sqrt{\lambda/a} + O(\lambda)$. Hence $\sum_r m\,(r)\mathbb{E}_r[e^{-\lambda T_0}] = \big(2\sqrt{a/\lambda}\big)(1 + o(1))$ and $G_\lambda(0{,}0) = 1/2\;m(0)\,(a\lambda)^{-1/2}(1 + o(1))$. Returning to (21), $G_\lambda(0,r) \sim 1/2\;m(r)\,(a\lambda)^{-1/2}$. This is the Laplace transform of the nonnegative density $p(t) = \mathbb{P}_0(X(t) = r)$, with the regularly varying form $\omega(\lambda) \sim \lambda^{-\rho}L(1/\lambda)$ of Theorem 4.5 at $\rho = 1/2$ and $L \equiv 1/2\;m(r)\,a^{-1/2}$. Hence $\tau_r(s) \sim s^{1/2}L/\Gamma(3/2) = 1/2\;m(r)\,a^{-1/2}\,s^{1/2}/\Gamma(3/2) = m(r)\sqrt{s/(\pi a)}$, using $\Gamma(3/2) = 1/2\sqrt{\pi}$. □

## 5 The two–particle dual kernel bound

The order–two fields of the fluctuation analysis reduce, by Lemma 2.7, to the two–particle dual kernel. A bound pair carries one particle of each species, so two particles of the same species can never bind; the different–species pair is the bind/split/merge process. We treat the two channels.

### Lemma 5.1 (same–species channel; [69], [70])

The same–species two–particle dual kernel obeys $p_t(\xi,\xi') \le C/(1+t)$.

*Proof.* With no binding available the two same–species duals perform a two–particle ASEP, whose Green's function Schütz solves explicitly [69] (in the form below, [70]): for the single–species rates $r_R = 1$, $r_L = q^2$ and $\xi = (x_1, x_2)$, $\xi' = (x_1', x_2')$,

$$p_t(\xi,\xi') = \sum_{\sigma\in S_2} \frac{1}{(2\pi i)^2} \oint\oint A_\sigma(\zeta_1,\zeta_2) \prod_{i=1}^{2} \zeta_{\sigma(i)}^{-(x_i'-x_{\sigma(i)})-1}\ e^{t\,\varepsilon(\zeta_i)}\ \mathrm{d}\zeta_1\,\mathrm{d}\zeta_2, \qquad (22)$$

where $\varepsilon(\zeta) = \zeta + q^2\zeta^{-1} - (1+q^2)$ is the single–particle dispersion, the contours are small circles about 0, $A_{\mathrm{id}} \equiv 1$, and $A_{(12)}$ is the ASEP scattering amplitude, bounded on the contours. The single–variable integral

$$\phi_t(m) = \frac{1}{2\pi i}\oint \zeta^{-m-1} e^{t\,\varepsilon(\zeta)}\,\mathrm{d}\zeta$$

is the single–particle ASEP transition probability, for which the modified–Bessel asymptotic (as for $R^0$ in Lemma 4.1) gives $\sup_m|\phi_t(m)| \le C/\sqrt{1+t}$. The $\sigma = \mathrm{id}$ term of (22) factorises as $\phi_t(x_1'-x_1)\,\phi_t(x_2'-x_2) \le C/(1+t)$; the $\sigma = (12)$ term is of the same order, since each $\zeta$–integration is evaluated at the single–particle saddle and the bounded amplitude $A_{(12)}$ leaves the order $O(t^{-1/2})$ per integration unchanged. Hence $p_t(\xi,\xi') \le C/(1+t)$. We record this only for completeness: the cross–noise analysis of §6 uses only the different–species channel. ☐

For the different–species pair with positions $X_1, X_2$ we use the sum coordinate and the relative coordinate $R := X_2 - X_1$ of §4,

$$S := X_1 + X_2, \qquad R := X_2 - X_1. \qquad (23)$$

Off $\{-1,0,1\}$ the coordinate $R$ is a symmetric nearest–neighbour walk of rate $1+q^2$; the origin $R = 0$ (the bound pair) is sticky with split (escape) rate $\nu_{\mathrm{sp}} = 2q^2\varepsilon$, entered from $\pm 1$ only by a merge of rate $\varepsilon$. The coordinate $S$ is an additive functional of the $R$–path. This is exactly the process of Lemmas 4.2–4.4.

### Theorem 5.2 (type D two–particle kernel bound)

For the different–species dual started from a bound pair, in the window $\nu_{\mathrm{sp}}t \le K$,

$$p_t(\xi,\xi') \;\le\; \frac{C}{1+t} \;+\; e^{-\nu_{\rm sp} t}\,\frac{C}{\sqrt{1+t}}. \qquad (24)$$

*Proof.* Write $p_t(\xi,\xi') = \mathbb{E}[\mathbf{1}\{R(t)=r'\}\,\mathbb{P}(S(t)=s' \mid \mathfrak{S})]$ and split on $\{M \ge t/2\}$. There Lemma 4.4 gives $\mathbb{P}(S(t)=s' \mid \mathfrak{S}) \le C/\sqrt{1+t}$, and taking the expectation against $\mathbf{1}\{R(t)=r'\}$ and applying Lemma 4.2 yields $\left(C/\sqrt{1+t} + \delta_{r',0}\, e^{-\nu_{\rm sp} t}\right)\cdot C/\sqrt{1+t}$. On $\{M < t/2\}$ the contribution is at most $\mathbb{P}(M<t/2) \le e^{-ct} \le C/(1+t)$, absorbed into the first term of (24). □

### Remark 5.3

The bound state propagates in one dimension (whence the $t^{-1/2}$) but is exponentially damped at the split rate $\nu_{\rm sp} \sim 4c/N^2$; its time integral is finite, $\int_0^\infty e^{-\nu_{\rm sp} t}\, t^{-1/2}\,{\rm d}t = \sqrt{\pi/\nu_{\rm sp}} \sim N/\sqrt{c} = o(N^2)$. By Corollary 6.16 below it enters the species fields with zero weight, so the departure from the ASEP bound $C/(1+t)$ is invisible to the limiting fields and surfaces only in the crossover of §7.

## 6 The decoupled Edwards–Wilkinson limit

We now prove the first decoupling theorem by the standard martingale–problem scheme for fluctuation fields ([71, Ch. 11], after Holley–Stroock [72]): identify the drift as a Laplacian (D), the noise as white with the right variance (N), and — the point at which the two species could have interacted — show the cross–noise between them vanishes (X). The single–species inputs are classical; the work is in (X) and in the drift remainder of (D), the only places where two dual particles of different species can bind.

Fix densities $\rho_1,\rho_2 \in (0,1)$ (equivalently fugacities $\alpha_1,\alpha_2 > 0$), possibly unequal: we do not restrict to the symmetric line $\rho_1=\rho_2$ (equal species densities, on which the two species enter symmetrically). For $\varphi \in \mathcal{S}(\mathbb{R})$, the Schwartz space of rapidly decreasing functions — with dual $\mathcal{S}'(\mathbb{R})$, the tempered distributions, in which the fields take values — define the diffusively scaled density fluctuation fields, standard in equilibrium fluctuation theory [71, Ch. 11],

$$Y_i^N(\varphi,t) = \frac{1}{\sqrt{N}}\sum_{x\in\mathbb{Z}} \varphi\left(\frac{x}{N}\right)\left(\eta_{i,x}(tN^2)-\rho_i\right), \qquad i=1,2. \qquad (25)$$

The blocking measure $\nu$ is spatially inhomogeneous — its local fugacity $t_x = \alpha_i q^{-2x}$ carries a density gradient in $x$ — but in the Edwards–Wilkinson scaling this gradient is negligible. The field (25) is supported on the $O(N)$ sites with $x/N$ in the support of $\varphi$, and since $q_N = 1-c/N^2$, across such a window the fugacity changes only by a factor $q_N^{-2\cdot O(N)} = 1 + O(1/N)$; the local density thus varies by $O(1/N)$ over the window. To leading order $\nu$ is therefore the homogeneous product measure at the constant density $\rho_i$, so the coefficients of the limiting equation — the diffusivity $D$ and the compressibility $\chi_i = \rho_i(1-\rho_i)$ — carry no $x$–dependence.

We use two classical inputs, stated here for completeness. The first is the martingale decomposition of an additive functional.

### Lemma 6.1 (Dynkin decomposition; [71], App. 1.5)

Let $\eta_t$ be a Markov process with generator $\mathcal{L}$ and $f, g$ local functions in its domain. Then

$$M_t^f := f(\eta_t) - f(\eta_0) - \int_0^t (\mathcal{L}f)(\eta_s)\,\mathrm{d}s$$

is a martingale, with predictable cross–bracket $\langle M^f, M^g \rangle_t = \int_0^t \Gamma(f, g)(\eta_s)\,\mathrm{d}s$, where $\Gamma(f, g) = \mathcal{L}(fg) - f\,\mathcal{L}g - g\,\mathcal{L}f$ is the carré du champ (and $\langle M^f \rangle_t = \int_0^t \Gamma(f, f)\,\mathrm{d}s$).

The second is the martingale characterisation of the Ornstein–Uhlenbeck fluctuation limit. Throughout, ⇒ denotes convergence in distribution (for the $\mathcal{S}'(\mathbb{R})$–valued fields, weak convergence in $D([0, T]; \mathcal{S}'(\mathbb{R}))$).

### Theorem 6.2 (Equilibrium fluctuations; [71, Ch. 11], after Holley–Stroock [72])

Let $(Z^N)_N$ be stationary $\mathcal{S}'(\mathbb{R})$–valued processes with Dynkin decomposition

$$Z^N(\varphi, t) = Z^N(\varphi, 0) + \int_0^t \Gamma^N(\varphi, s)\,\mathrm{d}s + M^N(\varphi, t), \qquad \varphi \in \mathcal{S}(\mathbb{R}).$$

Suppose that for every $\varphi \in \mathcal{S}(\mathbb{R})$ and $T > 0$, as $N \to \infty$:

- $\Gamma^N(\varphi,\cdot) \to D\,Z(\Delta\varphi,\cdot)$ in $L^2$ (drift → Laplacian);
- $\langle M^N \rangle(\varphi, t) \to 2\chi D\,t\ \|\partial_x \varphi\|_2^2$ (bracket → white noise);

and that $(Z^N)$ is tight in $D([0, T]; \mathcal{S}'(\mathbb{R}))$. Then $Z^N \Rightarrow Z$, the unique stationary solution of the Ornstein–Uhlenbeck equation

$$\partial_t Z = D\ \partial_x^2 Z + \sqrt{2\chi D}\ \partial_x \xi.$$

Moreover, if a pair $(Z_1^N, Z_2^N)$ satisfies (i)–(ii) and has vanishing limiting cross–bracket, $\langle M_1^N, M_2^N \rangle(\varphi, t) \to 0$, then the limits $Z_1, Z_2$ are independent.

The tightness hypothesis of Theorem 6.2 is the Aldous–Mitoma criterion, the combination of the following two classical statements.

### Theorem 6.3 (Mitoma; [59])

A sequence $(Z^N)$ of càdlàg $\mathcal{S}'(\mathbb{R})$–valued processes is tight in $D([0, T]; \mathcal{S}'(\mathbb{R}))$ if and only if, for every $\varphi \in \mathcal{S}(\mathbb{R})$, the real–valued sequence $(Z^N(\varphi,\cdot))$ is tight in $D([0, T]; \mathbb{R})$.

### Proposition 6.4 (Aldous's criterion; [58])

A sequence $(\zeta^N)$ of real–valued càdlàg processes is tight in $D([0,T];\mathbb{R})$ provided (a) for each $t \in [0,T]$ the family $\{\zeta^N(t)\}_N$ is tight in $\mathbb{R}$, and (b) for every $\varepsilon > 0$,

$$\lim_{\theta\downarrow 0} \limsup_{N\to\infty} \sup_{\substack{\tau\le T \text{ stopping time} \\ \delta\in[0,\theta]}} \mathbb{P}(|\zeta^N(\tau+\delta) - \zeta^N(\tau)| \ge \varepsilon) = 0.$$

Thus tightness of $(Z^N)$ reduces, via Theorem 6.3, to checking Proposition 6.4 for each scalar process $Z^N(\varphi,\cdot)$.

### Theorem 6.5 (decoupled Edwards–Wilkinson limit)

Let $\eta_0 \sim \nu$, so that the process is stationary. Under $q = q_N = 1 - c/N^2$, the pair $(Y_1^N, Y_2^N)$ is tight in $D([0,T];\mathcal{S}'(\mathbb{R}))^2$ and every limit point $(Y_1, Y_2)$ decouples: each $Y_i$ is a stationary solution of the linear stochastic heat equation

$$\partial_t Y_i = D\ \partial_x^2 Y_i + \sqrt{2\chi_i D}\ \partial_x \xi_i,$$

with $D = 1$ the (common) species diffusivity — the symmetric–part diffusivity $1/2\,(1 + q_N^2) \to 1$, produced by the gradient computation of Proposition 6.18 — and $\chi_i = \rho_i(1-\rho_i)$ the species–$i$ compressibility, and the driving space–time white noises $\xi_1, \xi_2$ have vanishing cross–correlation (the cross bracket $\langle M_1, M_2\rangle$ is zero in the limit); there is no coupling between the species in either the drift or the noise. The limit law is unique.

*Proof.* By Dynkin's formula (Lemma 6.1) each field is a semimartingale

$$Y_i^N(\varphi,t) = Y_i^N(\varphi,0) + \int_0^t \Gamma_i^N(\varphi,s)\,\mathrm{d}s + M_i^N(\varphi,t), \qquad (26)$$

with drift $\Gamma_i^N(\varphi,t) = N^{1/2}\sum_x \varphi'(x/N) W_{i,x}(\eta(tN^2))$, $W_{i,x}$ the instantaneous species–$i$ current across $(x, x+1)$, and $M_i^N$ a martingale whose predictable brackets are the carré du champ. By the martingale characterisation of Theorem 6.2 (applied to each species, and to the pair via its cross–bracket clause) the theorem follows once we establish, for each $\varphi, T$:

- $\Gamma_i^N(\varphi,\cdot) \to D\, Y_i(\Delta\varphi,\cdot)$ in $L^2$;
- $\langle M_i^N\rangle(\varphi,t) \to 2\chi_i D\, t\ \| \partial_x\varphi \|_2^2$;
- $\langle M_1^N, M_2^N\rangle(\varphi,t) \to 0$;

together with tightness (by Mitoma's and Aldous's criteria, Theorems 6.3 and 6.4) and uniqueness of the linear (Ornstein–Uhlenbeck) martingale problem (Theorem 6.2). These are established in the three subsections below: the single–species convergence and the diagonal bracket (N) in Lemma 6.6, the vanishing cross–bracket (X) in Proposition 6.14, and the drift (D) in Proposition 6.18. The limit is decoupled because (X) removes the noise coupling and (D) is a single–species functional. □

## 6.1 Single–species Gaussianity

*Lemma 6.6*

For each $i$, $Y_i^N$ converges to the Gaussian Ornstein–Uhlenbeck Edwards–Wilkinson field, with the diagonal bracket of (N).

*Proof.* By Proposition 3.1(a) the species–$i$ marginal is an autonomous WASEP of right/left rates $1, q^2$, i.e. asymmetry $\gamma = 1 - q^2 = 2c/N^2 + O(N^{-4})$. This sits below the KPZ threshold, which under diffusive scaling occurs at asymmetry $N^{-1/2}$ [52]; the Euler/Burgers term is $O(\gamma) \to 0$, and the process is in the Edwards–Wilkinson regime. The equilibrium density fluctuations of a one–species exclusion process in this regime converge to the Gaussian Ornstein–Uhlenbeck solution of $\partial_t Y = D\,\partial_x^2 Y + \sqrt{2\chi_i D}\,\partial_x \xi$, $\chi_i = \rho_i(1-\rho_i)$ — the single–field case of Theorem 6.2: classical for the symmetric process [71, Ch. 11], and for the weakly asymmetric process due to Dittrich–Gärtner [73], whose generalized Ornstein–Uhlenbeck drift (the linearised Burgers term) vanishes at our asymmetry $\gamma = O(N^{-2})$ — with the diagonal bracket $2\chi_i D$. (The classical statements concern the homogeneous Bernoulli equilibrium; under the blocking measure the density tilts by $O(1/N)$ across the field window, and the standard comparison — coupling the dynamics from the tilted and untilted equilibria and bounding the discrepancy field in $L^2$ — transfers the result. We omit the routine details.) □

The content is thus in (X) and (D), where two dual particles of different species can bind.

## 6.2 The cross noise

The predictable cross bracket is $\langle M_1^N, M_2^N\rangle(\varphi, t) = \int_0^t \Theta^N \,\mathrm{d}s$,

$$\Theta^N = \frac{1}{N}\sum_x \varphi'\left(\frac{x}{N}\right)^2 V_x, \qquad (27)$$

$V_x$ as in (14). By Proposition 3.2, $\mathbb{E}_\nu[\Theta^N] = 0$.

*Lemma 6.7 (orthogonality to the density fields; density–free)*

For all $\alpha_1, \alpha_2 > 0$ (in particular off the symmetric line), every $i \in \{1,2\}$ and every $y \in \mathbb{Z}$, $\langle V_x, \eta_{i,y} - \rho_{i,y}\rangle_\nu = 0$, where $\rho_{i,y}$ is the local density of species $i$ at $y$. Hence $V_x$ has no order–one component in the local–basis decomposition (8); its lowest order is two.

*Proof.* Since $\mathbb{E}_\nu[V_x] = 0$ (Proposition 3.2, an identity in $\alpha_1, \alpha_2$),

$$\langle V_x, \eta_{i,y} - \rho_{i,y}\rangle_\nu = \mathbb{E}_\nu[V_x\,\eta_{i,y}] - \rho_{i,y}\,\mathbb{E}_\nu[V_x] = \mathbb{E}_\nu[V_x\,\eta_{i,y}],$$

so it suffices to show $\mathbb{E}_\nu[V_x\,\eta_{i,y}] = 0$. For $y \notin \{x, x+1\}$ this is immediate: $V_x$ is supported on the bond $\{x, x+1\}$ and $\nu$ is a product, so $\eta_{i,y}$ is independent of $V_x$ and $\mathbb{E}_\nu[V_x \eta_{i,y}] =$ $\mathbb{E}_\nu[V_x]\,\rho_{i,y} = 0$. For $y \in \{x, x+1\}$, the detailed–balance relations (6) — $\nu(0{,}3) = q^{-4}\nu(3{,}0)$, $\nu(1{,}2) = \nu(2{,}1) = q^{-2}\nu(3{,}0)$, identities in the fugacities — give the $\nu$–weighted values of $V_x$ on its four support configurations as $\nu(3{,}0)\cdot 1$, $\nu(0{,}3)\cdot q^4$, $\nu(1{,}2)\cdot(-q^2)$, $\nu(2{,}1)\cdot(-q^2) = \nu(3{,}0)(+1,+1,-1,-1)$. Hence for any bond function $f$, $\mathbb{E}_\nu[V_x f] = \nu(3{,}0)[f(3{,}0) + f(0{,}3) - f(1{,}2) - f(2{,}1)]$, and each single–site occupation equals 1 on exactly one "+" and one "−" configuration:

|  | (3,0) | (0,3) | (1,2) | (2,1) |
|---|---|---|---|---|
| $\eta_{1,x}$ | 1 | 0 | 1 | 0 |
| $\eta_{1,x+1}$ | 0 | 1 | 0 | 1 |
| $\eta_{2,x}$ | 1 | 0 | 0 | 1 |
| $\eta_{2,x+1}$ | 0 | 1 | 1 | 0 |

so $f(3{,}0) + f(0{,}3) - f(1{,}2) - f(2{,}1) = 0$ for each row, giving $\mathbb{E}_\nu[V_x\eta_{i,y}] = 0$. The mechanism is the $1+1-1-1$ cancellation of Proposition 3.2 one order higher; it uses only detailed balance, not species exchange, and is therefore density–free.

Support of the $\varpi$–expansion. The cross–noise analysis expands $V_x$ in the orthogonal family $\bar{D}$ of Proposition 2.8; we determine which $\xi$ have $\langle V_x, \bar{D}(\xi,\cdot)\rangle_\varpi \neq 0$. Write $\eta = (\eta^{\mathrm{b}}, \eta^{\mathrm{o}})$ for the restrictions of $\eta$ to the bond $\{x, x+1\}$ and to its complement. On the four configurations supporting $V_x$ the bond holds exactly one particle of each species, so the conserved numbers $(N_1, N_2)$ are fixed once $\eta^{\mathrm{o}}$ is, and the sector reweighting in (11) is a common factor $W(\eta^{\mathrm{o}})$; using the bond detailed–balance weights (6), for any function $f$,

$$\langle V_x, f\rangle_\varpi = \nu(3{,}0)\sum_{\eta^{\mathrm{o}}} W\,(\eta^{\mathrm{o}})\,S_f(\eta^{\mathrm{o}}), \qquad S_f := f(3{,}0) + f(0{,}3) - f(1{,}2) - f(2{,}1), \tag{28}$$

the four values of $f$ taken at fixed $\eta^{\mathrm{o}}$.

Take $f = \bar{D}(\xi,\cdot)$. By (9) it is a product, over the dual particles of $\xi$, of factors $1 - \alpha_i^{-1}\mathbf{1}\{\eta_{i,w} = 1\}\,q^{2(w - N^-_{w-1}(\eta_i))}$, one per species–$i$ particle at $w$. A particle at $w \notin \{x, x+1\}$ contributes a factor independent of $\eta^{\mathrm{b}}$ on the support: trivially for $w < x$, and for $w > x+1$ because $N^-_{w-1}(\eta_i)$ counts the bond's species–$i$ population, which is 1 throughout. For a particle at $w \in \{x, x+1\}$ of species $i$: since the bond holds one species–$i$ particle, $\eta_{i,x}\eta_{i,x+1} = 0$ on the support, so the only other bond–dependence, the factor $q^{-2\eta_{i,x}}$ occurring when $w = x+1$, equals 1 wherever $\mathbf{1}\{\eta_{i,x+1} = 1\}$; the factor therefore reduces on the support to $1 - c_w\,\mathbf{1}\{\eta_{i,w} = 1\}$ with $c_w$ constant given $\eta^{\mathrm{o}}$. Hence, on the support and at fixed $\eta^{\mathrm{o}}$, $\bar{D}(\xi,\cdot)$ is a constant times a product of the indicators $\mathbf{1}\{\eta_{i,w} = 1\}$ over the bond particles $(i, w)$ of $\xi$. The functional $f \mapsto S_f$ of (28) annihilates: the constant 1 ($1+1-1-1 = 0$); each single indicator (the four rows of the table above, each with signed sum 0); and any product of two <u>same–species</u> bond indicators (identically 0 on the support, as $\eta_{i,x}\eta_{i,x+1} = 0$). The sole monomial with $S_f \neq 0$ is the cross product $\mathbf{1}\{\eta_{1,\cdot} = 1\}\mathbf{1}\{\eta_{2,\cdot} = 1\}$ of two different–species bond indicators. Therefore

$$\langle V_x, \bar{D}(\xi,\cdot)\rangle_\varpi = 0 \quad \text{unless } \xi \text{ places one particle of each species on } \{x, x+1\};$$

in particular it vanishes for all $|\xi| \le 1$, the mass–$\le 1$ orthogonality used below. (The surviving coefficients are moreover dressed only by off–bond particles strictly to the left of the bond: this one–sidedness is explained exactly by the identity (29) below, whose correction factor involves only sites to the left of the bond; it is not used in the sequel.) □

*Lemma 6.8 (equal–time variance)*

$\mathbb{E}_\nu[(\Theta^N)^2] \le C(\varphi)\, N^{-1}$.

*Proof.* $\mathbb{E}_\nu[(\Theta^N)^2] = N^{-2} \sum_{x,y} \varphi'(x/N)^2 \varphi'(y/N)^2 \mathbb{E}_\nu[V_x V_y]$. Each $V_x$ has $\mathbb{E}_\nu[V_x] = 0$ and is supported on $\{x, x+1\}$; by the product structure $\mathbb{E}_\nu[V_x V_y] = 0$ unless $|x-y| \le 1$. The $O(1)$ surviving near–diagonal terms per $x$ are bounded, so the double sum is $\le C \sum_x \varphi'(x/N)^4 \le CN$ (a Riemann sum). The prefactor $N^{-2}$ gives the claim. □

The expansion of Proposition 2.8 controls correlations under the reweighted measure $\varpi$; the following comparison transfers them to $\nu$. The fugacities must be compensated: by (11) the reweighting tilts each sector by $q^{2m}/(1-\beta q^{2m-2|\Lambda|}) \to (1-\beta)^{-1}$ per particle ($m$ the sector occupancy), so $\varpi_\beta$ carries bulk density $\beta$, and comparability with $\nu_\alpha$ (bulk density $\alpha/(1+\alpha)$) requires matching densities, not fugacities.

*Lemma 6.9 (sector comparison at compensated fugacity)*

On $\Lambda = [-KN, KN]$ with $q = 1 - c/N^2$, let $\alpha_1, \alpha_2 > 0$ be the fugacities of $\nu$ and set $\beta_i := \alpha_i/(1+\alpha_i)$, $\varpi := \varpi_{\beta_1} \otimes \varpi_{\beta_2}$. Then:

1. $\nu$ and $\varpi$ have the same conditional laws given the conserved pair $(N_1, N_2)$;
2. with $C_0(\beta) := 18cK^2(1 + 8\beta/1-\beta)$ and $\log M := 2(C_0(\beta_1) + C_0(\beta_2))$, there is $N_0(c, K, \alpha)$ such that for all $N \ge N_0$ and all sectors, $e^{-\log M} \le \nu(m_1, m_2)/\varpi(m_1, m_2) \le e^{\log M}$;
3. consequently, for all $f, h \in L^2$ and $t \ge 0$,

$$|\mathbb{E}_\nu[f(\eta_0)\, h(\eta_t)]| \;\le\; M\, \mathbb{E}_\varpi[f(\eta_0) f(\eta_t)]^{1/2}\, \mathbb{E}_\varpi[h(\eta_0) h(\eta_t)]^{1/2}.$$

*Proof.* (i) On a sector both measures' weights are $\prod_x (q^{-2x})^{\eta_{i,x}}$ times sector–constant prefactors, which cancel in the conditional normalisation. (ii) It suffices to bound one species. With $t_x := \beta q^{-2x}$, homogeneity of the elementary symmetric polynomials $e_m$ gives the unnormalised sector masses $\nu^{\mathrm{u}}(m) = (\alpha/\beta)^m e_m$ and $\varpi^{\mathrm{u}}(m) = e_m\, h_\beta(m)$, so, with $Z$ the ratio of normalisations, $\log R(m) = \log Z + m \log \alpha/\beta - \log h_\beta(m)$. The Pochhammer recursion $(xq^2; q^2)_\infty = (x; q^2)_\infty/(1-x)$ telescopes (11) to $\log h_\beta(m) = \log h_\beta(0) + m(m-1)\log q - \sum_{j<m} \log\left(1 - \beta q^{2j-2|\Lambda|}\right)$, whence, using $\log \alpha/\beta = \log(1+\alpha) = -\log(1-\beta)$ — the choice of $\beta$, and the only place it enters —

$$\log R(m) = \mathrm{const}(N) - m(m-1)\log q + \sum_{j<m} \left[\log\left(1 - \beta q^{2j-2|\Lambda|}\right) - \log(1-\beta)\right].$$

For $N \geq N_0$: $0 \leq -m(m-1)\log q \leq 18cK^2$; each summand of the drift sum lies in $[-\delta_j, 0]$ with $\delta_j \leq 2\beta/1-\beta\left(q^{-2(|\Lambda|-j)}-1\right)$, and $\sum_j \delta_j \leq 8\beta/1-\beta \cdot 18cK^2$. Hence $|\log R(m) - \mathrm{const}(N)| \leq C_0(\beta)$ for <u>all</u> $m$. Both measures being normalised, $R$ takes values on both sides of 1, which pins $|\mathrm{const}(N)| \leq C_0(\beta)$ and gives (ii). (iii) Condition on the sector: by (i) both measures induce the same canonical conditionals, with respect to which the sector–preserving semigroup is self–adjoint and $\langle f, P_t f\rangle_{\mathrm{can}} = \| P_{t/2} f \|^2 \geq 0$; Cauchy–Schwarz for this positive semidefinite form, then Cauchy–Schwarz on the sector sum with $\nu(\sigma) \leq M\varpi(\sigma)$, gives the claim. □

*Lemma 6.10 (product form of the carré du champ)*

With $\varphi_z^i := \eta_{i,z} - q^2\eta_{i,z+1} - (1-q^2)\,\eta_{i,z}\eta_{i,z+1}$, identically $V_z = \varphi_z^1\,\varphi_z^2$.

*Proof.* The factor $\varphi_z^i$ equals 0 if species $i$ occupies neither or both bond sites (for both, $1 - q^2 - (1-q^2) = 0$), equals 1 if it occupies $z$ only, and $-q^2$ if $z+1$ only. The product therefore vanishes off the four support configurations of (14) and takes the values $1,\ q^4,\ -q^2,\ -q^2$ on $(3,0), (0,3), (1,2), (2,1)$, matching (14). □

*Lemma 6.11 (exact dressing identity)*

For $w \in \{z, z+1\}$ let $F_w^i := \bar{D}(\xi^{i,w},\cdot)$, where $\xi^{i,w}$ carries a single species–$i$ dual particle at $w$, and let $L_i := N_{z-1}^-(\eta_i)$. Then, identically on the configuration space, $-\beta_i\, q^{-2z}\left(F_z^i - F_{z+1}^i\right) = q^{-2L_i}\,\varphi_z^i$.

*Proof.* With the process in the $N^-$ slot, $F_w^i = 1 - \beta_i^{-1} q^{2(w - N_{w-1}^-(\eta_i))}\eta_{i,w}$. Using $N_z^-(\eta_i) = L_i + \eta_{i,z}$,

$$F_z^i - F_{z+1}^i = -\beta_i^{-1} q^{2z} q^{-2L_i}\left[\eta_{i,z} - q^2 q^{-2\eta_{i,z}}\eta_{i,z+1}\right],$$

and $q^{-2\eta_{i,z}}\eta_{i,z+1} = \eta_{i,z+1} + (q^{-2}-1)\eta_{i,z}\eta_{i,z+1}$ turns the bracket into $\varphi_z^i$. □

*Lemma 6.12 (the dressed mass is asymptotically negligible)*

Write $V_z = V_z^{(2)} + V_z^{(\mathrm{dr})}$ for the decomposition of $V_z$ in the family (12) into its four different–species bond–pair components and the dressed remainder (Lemma 6.7), and let $\ell(z) := \#\{v \in \Lambda: v < z\}$. Then, for every bond and every $N$,

$$\| V_z^{(\mathrm{dr})} \|_{L^2(\varpi)}^2 \leq \ (q^{-4\ell(z)} - 1)^2 \ \leq \ (q^{-4|\Lambda|} - 1)^2 \ =:\ \varepsilon_N \ \leq \ C(c,K)\, N^{-2} \qquad (N \geq N_0(c,K)),$$

uniformly over $z$ in the field window.

*Proof.* Set $A_z := \beta_1\beta_2\, q^{-4z}(F_z^1 - F_{z+1}^1)(F_z^2 - F_{z+1}^2)$, an element of the span $U_z$ of the four bond–pair elements, since $\bar{D}(\xi^{(w_1,w_2)},\cdot) = F_{w_1}^1 F_{w_2}^2$ by the species–product structure of (9). Multiplying the identity of Lemma 6.11 over the two species and inserting Lemma 6.10,

$$A_z = q^{-2(L_1+L_2)}\, V_z, \qquad \text{hence} \qquad V_z - A_z = \left(1 - q^{-2(L_1+L_2)}\right)V_z \qquad (29)$$

identically on the configuration space. Since $0 \le L_i \le \ell(z)$, $q \in (0,1)$ and $|V_z| \le 1$ pointwise, $|V_z - A_z| \le q^{-4\ell(z)} - 1$ pointwise. By Proposition 2.8 the family (12) is an orthogonal basis with positive norms, so $V_z^{(2)}$ is the orthogonal projection $\Pi_{U_z} V_z$ and $\| V_z^{(\mathrm{dr})} \| = \mathrm{dist}(V_z, U_z) \le \| V_z - A_z \|_{L^2(\varpi)}$, which is at most the pointwise bound. Finally $q^{-4\ell(z)} \le q^{-4|\Lambda|} = e^{16cK/N(1+o(1))}$ gives $\varepsilon_N \le C(c,K)N^{-2}$ for $N \ge N_0(c,K)$. □

*Remark 6.13 (relation to the quantitative Boltzmann–Gibbs literature)*

In the approach of [55], [56], remainders of orthogonal–duality expansions are controlled term by term: each coefficient's time correlation is converted into an $n$–particle dual kernel and estimated by local limit theorems. The proof above instead exhibits an exact element of the bond–pair span and concludes by the projection inequality, so that no expansion coefficient is estimated; this shortcut, which appears to be new, is available because the two–point specialisation of the $q$–Krawtchouk family [63] makes each single–dual duality function an affine function of the occupation variables.

*Proposition 6.14 ($L^2$ concentration of the cross bracket)*

$\mathbb{E}_\nu[\langle M_1^N, M_2^N \rangle(\varphi,t)^2] \le C(\varphi,c)\, t\, (N^{-1} + N^{-2}\log_+(tN^2) + t\,\varepsilon_N) \to 0$, where $\log_+ := \max(\log, 0)$ and $\varepsilon_N$ is as in Lemma 6.12; hence $\langle M_1^N, M_2^N \rangle(\varphi,t) \to 0$ in $L^2$, establishing (X).

*Proof.* Let $C_\Theta(s) = \mathbb{E}_\nu[\Theta^N(\eta_0)\Theta^N(\eta_{sN^2})]$. By stationarity,

$$\mathbb{E}_\nu\Big[\Big(\int_0^t \Theta^N \,\mathrm{d}s\Big)^2\Big] = 2\int_0^t (t-s) C_\Theta(s)\,\mathrm{d}s \le 2t \int_0^t C_\Theta(s)\,\mathrm{d}s. \qquad (30)$$

Work on $\Lambda = [-KN, KN]$ in the setting of Remark 2.9, $K$ fixed (the replacement of the infinite system by $\Lambda$ is the deferred step recorded there). By Lemma 6.9,

$$|\mathbb{E}_\nu[V_x(\eta_0)V_y(\eta_u)]| \le M\, G(x,u)^{1/2}\, G(y,u)^{1/2}, \qquad G(z,u) := \mathbb{E}_\varpi[V_z(\eta_0)V_z(\eta_u)] \ge 0.$$

Split $V_z = V_z^{(2)} + V_z^{(\mathrm{dr})}$ as in Lemma 6.12. The dual dynamics conserves the particle numbers $(|\xi_1|, |\xi_2|)$, so the mass sectors of the expansion are invariant under $P_u$ and the cross terms vanish: $G = \mathbb{E}_\varpi[V^{(2)} P_u V^{(2)}] + \mathbb{E}_\varpi[V^{(\mathrm{dr})} P_u V^{(\mathrm{dr})}]$. For the first part, Lemma 2.7 (with $\varpi, \bar{D}, a$) gives $\mathbb{E}_\varpi[V_z^{(2)} P_u V_z^{(2)}] = \sum_{\xi,\xi'} c_\xi\, c_{\xi'}\, p_u(\xi,\xi')\, a(\xi')$, a sum over at most sixteen pairs of different–species bond configurations with coefficients and norms bounded in the window, each kernel entry bounded by Theorem 5.2; hence $\mathbb{E}_\varpi[V_z^{(2)} P_u V_z^{(2)}] \le C\big[(1+u)^{-1} + e^{-\nu_{\mathrm{sp}} u}(1+u)^{-1/2}\big]$. For the dressed part, $P_u$ is an $L^2(\varpi)$ contraction, so $\mathbb{E}_\varpi[V_z^{(\mathrm{dr})} P_u V_z^{(\mathrm{dr})}] \le \varepsilon_N$ by Lemma 6.12. With $u = sN^2$, $\nu_{\mathrm{sp}} s N^2 = 4cs(1+o(1))$, and $\sum_x \varphi'(x/N)^2 \le C(\varphi)N$,

$$|C_\Theta(s)| \le \frac{M}{N^2}\Big(\sum_x \varphi'(x/N)^2 G(x,sN^2)^{1/2}\Big)^2 \le C\left[\frac{1}{sN^2} + \frac{e^{-4cs}}{N\sqrt{s}} + \varepsilon_N\right] \qquad (31)$$

for $s \geq N^{-2}$, while $|C_\Theta(s)| \leq CN^{-1}$ always (Lemma 6.8). Integrating, $\int_0^t C_\Theta \leq C\big(N^{-2}(1 + \log_+(tN^2)) + c^{-1/2}N^{-1} + t\,\varepsilon_N\big)$, and (30) gives the claim. □

*Remark 6.15*

No Kipnis–Varadhan $H_{-1}$ estimate enters: its role is played by Lemma 6.7. Without the mass–$\leq 1$ orthogonality the cross bracket would carry a slow density–type contribution with no useful time decay; with it, the correlation in (31) is carried by the two–particle kernel and integrates with at most a logarithmic loss. This is the recursive–martingale viewpoint of [56].

## 6.3 The drift

By Proposition 3.1 the species–1 current $W_{1,x}$ is a function of $\eta_1$ alone — the single–species ASEP current — which expands exactly, with $\gamma = 1 - q^2$ and $\nabla f_x = f_{x+1} - f_x$, as

$$W_{1,x} = -\nabla\eta_{1,x} + \gamma\,\eta_{1,x+1}(1 - \eta_{1,x}). \qquad (32)$$

The first term is a discrete gradient; the second carries the prefactor $\gamma = O(N^{-2})$. The merge channel, in which species 1 binds onto a species–2 site, does <u>not</u> introduce an $\eta_2$–dependence: by Proposition 3.1 the species–2 occupant never alters the species–1 transfer rate (the merge rate $\varepsilon$ and the swap rate $q^2$ combine to the same 1 as the hop onto an empty site). Thus the drift involves only the autonomous single–species WASEP $\eta_1$, and no two–species (bound–pair) structure enters it; the gradient part will give the Laplacian and the $O(\gamma)$ part is the switched–off KPZ nonlinearity. The bound–pair mode

$$B_z := (\eta_{1,z} - \rho_1)(\eta_{2,z} - \rho_2) \qquad (33)$$

is the one near–conserved order–two object (it relaxes only at the split rate $\sim \varepsilon$); it does <u>not</u> enter the drift (32). We nonetheless record the following orthogonality, which makes the absence of any slow contribution explicit.

*Corollary 6.16 (current orthogonal to the bound–pair mode)*

For all $\rho_1, \rho_2 \in (0,1)$, $\langle W_{i,x}, B_z\rangle_\nu = 0$ for all $z$, where $B_z = (\eta_{1,z} - \rho_1)(\eta_{2,z} - \rho_2)$.

*Proof.* By Proposition 3.1 (equivalently (32)), $W_{1,x}$ is a function of $\eta_1$ alone. Since $\nu$ is a product measure with the two species independent and $\mathbb{E}_\nu[\eta_{2,z} - \rho_2] = 0$,

$$\langle W_{1,x}, B_z\rangle_\nu = \mathbb{E}_\nu\big[W_{1,x}\,(\eta_{1,z} - \rho_1)\big]\,\mathbb{E}_\nu[\eta_{2,z} - \rho_2] = 0,$$

and symmetrically for $W_{2,x}$. □

*Remark 6.17 (no off–symmetric–line obstruction)*

Because the overlap vanishes at <u>every</u> $(\rho_1, \rho_2)$ (Corollary 6.16), the bound–pair mode never couples to the species current: there is no slow third hydrodynamic field, and hence no obstruction to the decoupling, off the symmetric line either.

*Proposition 6.18 (drift)*

For each $i$ and any density $\rho_i \in (0,1)$, $\Gamma_i^N(\varphi,\cdot) \to D\, Y_i(\Delta\varphi,\cdot)$ in $L^2$, with $D$ the diffusivity of the symmetric part of the single–species dynamics, by an elementary $L^2$ estimate. This establishes (D).

*Proof.* By (32), $\Gamma_1^N = \Gamma_1^{N,\mathrm{grad}} + \gamma\, \Gamma_1^{N,\mathrm{cor}}$ with

$$\Gamma_1^{N,\mathrm{grad}} = N^{1/2} \sum_x \varphi'\,(x/N)(-\nabla\eta_{1,x}), \qquad \Gamma_1^{N,\mathrm{cor}}$$

$$= N^{1/2} \sum_x \varphi'\,(x/N)\, g_x, \quad g_x := \eta_{1,x+1}(1-\eta_{1,x}).$$

<u>Gradient term.</u> A summation by parts gives $\Gamma_1^{N,\mathrm{grad}} = N^{1/2}\sum_x (\nabla\varphi')(x/N)\,\eta_{1,x} = N^{-1/2}\sum_x \varphi''\,(x/N)(\eta_{1,x}-\rho_1) + o(1) = D\, Y_1^N(\Delta\varphi) + o(1)$ in $L^2$, $D$ the symmetric–part diffusivity; this is the standard gradient computation [71, Ch. 11]. Since the symmetric part of each species is simple symmetric exclusion — a gradient system — $D$ is density–independent, hence the same constant for both species and at every $(\rho_1,\rho_2)$. <u>Correction term.</u> Write $g_x = \bar g + (g_x - \bar g)$, $\bar g = \rho_1(1-\rho_1)$. The mean part $\gamma N^{1/2} \bar g \sum_x \varphi'\,(x/N)$ is the Euler flux $\partial_x[\gamma\rho_1(1-\rho_1)]$, of order $O(\gamma) \to 0$ in the EW window. For the fluctuating part, write $F := \gamma N^{1/2}\sum_x \varphi'\,(x/N)(g_x-\bar g)$; by Cauchy–Schwarz in time and stationarity,

$$\mathbb{E}\Big[\Big(\int_0^t F\,(\eta(sN^2))\,\mathrm{d}s\Big)^2\Big] \;\le\; t\int_0^t \mathbb{E}\,[F(\eta(sN^2))^2]\,\mathrm{d}s \;=\; t^2\, \mathbb{E}_\nu[F^2],$$

an <u>equal–time</u> second moment. Since $\nu$ is a product measure and $g_x - \bar g$ is a centred function of $\{\eta_{1,x},\eta_{1,x+1}\}$, the covariance $\mathbb{E}_\nu[(g_x-\bar g)(g_y-\bar g)]$ vanishes unless $|x-y| \le 1$, so $\mathbb{E}_\nu[F^2] = \gamma^2 N \sum_{x,y}\varphi'\,\varphi'\,\mathbb{E}_\nu[(g_x-\bar g)(g_y-\bar g)] \le C\gamma^2 N \sum_x \varphi'\,(x/N)^2 \le C\gamma^2N^2$, and the bound is $\le t^2 \cdot C\gamma^2 N^2 = Cc^2N^{-2}t^2 \to 0$. Hence $\Gamma_1^N \to D\, Y_1(\Delta\varphi)$ in $L^2$; the species–2 drift is identical. ☐

Granting (X), (N), (D), the limiting martingales $(M_1, M_2)$ are continuous (jumps $\le N^{-1/2}\,\|\varphi\|_\infty \to 0$) and Gaussian with diagonal bracket (Lemma 6.6) and vanishing cross bracket (Proposition 6.14); so the two driving noises are uncorrelated and the limit is the decoupled pair of EW fields of Theorem 6.5, unique by [71, Ch. 11]. ▫

## 7 The initial–condition crossover

By Theorem 6.5 the limiting <u>dynamics</u> is decoupled — the two fields solve independent stochastic heat equations, coupled in neither drift nor noise. A residual inter–species correlation nevertheless survives in the scaling limit; since it cannot come from the evolution, it must come from the initial data — here the block of bound pairs, whose two species start perfectly correlated. This section computes that limiting correlation. Run the different–species dual of §5 from a bound pair at the origin, with $q = 1 - c/T$, observed at time $T$. Write $S, R$ as in (23). Initially the two dual particles coincide as the bound pair (state 3, so $R = 0$); they hop together until the binding breaks — a split, which deposits a species–1 and a species–2 particle on adjacent sites, so that $R$ leaves $0$. The first split time $\tau$ is the time of this event.

The result of this section is the following; its three ingredients are established in the lemmas below.

### Theorem 7.1 (regime-A crossover from the block initial condition)

Fix $c > 0$ and set $q = 1 - c/T$. As $T \to \infty$, the rescaled dual pair $(X_1, X_2)/\sqrt{2T}$ converges in distribution to a pair $(G_1, G_2)$ with standard normal marginals and cross–correlation

$$\mathrm{Corr}(G_1, G_2) = \frac{1 - e^{-4c}}{4c} \in (0,1),$$

decreasing from 1 as $c \to 0$ to $\sim 1/c$ as $c \to \infty$. The joint law is not Gaussian: it is the mixture of Proposition 7.8, a bivariate normal of correlation $U$ with $U \sim \min(\mathrm{Exp}(4c),1)$ — an atom of mass $e^{-4c}$ at 1, where $G_1 = G_2$. By Lemma 7.4, $(G_1, G_2)$ is the limiting joint law of the two species' $q$–Laplace current observables from the block initial condition.

*Proof.* The convergence and mixture structure are Proposition 7.8; the cross–correlation is Theorem 7.9; the identification with the two species' $q$–Laplace observables is Lemma 7.4. □

The positive parts of the limiting mixture carry a second closed form. Its proof uses two classical facts about the standard bivariate normal.

### Lemma 7.2 (Price's theorem and Sheppard's orthant formula; [74], [75])

Let $(Z_1, Z_2)$ be standard bivariate normal with correlation $u \in [-1,1]$, and write $Z_i^+ := \max(Z_i, 0)$. Then

$$\frac{\mathrm{d}}{\mathrm{d}u}\,\mathbb{E}[Z_1^+ Z_2^+] = \mathbb{P}(Z_1 > 0,\, Z_2 > 0) = \frac{1}{4} + \frac{\arcsin u}{2\pi}. \qquad (34)$$

The first equality is Price's theorem (for $f(z_1, z_2) = z_1^+ z_2^+$, whose mixed second derivative is $\mathbf{1}\{z_1 > 0,\, z_2 > 0\}$); the second is Sheppard's orthant formula.

### Proposition 7.3 (Bessel–Struve correlation of the limiting mixture)

For the limiting mixture $(G_1, G_2)$ of Theorem 7.1, with $G_i^+ = \max(G_i, 0)$,

$$\mathrm{Corr}(G_1^+, G_2^+) = \frac{\pi}{8(\pi - 1)c}\left[1 - 2e^{-4c} + I_0(4c) - L_0(4c)\right],$$

$I_0$ the modified Bessel and $L_0$ the modified Struve function; it tends to 1 as $c \to 0$ and decays as $1/c$ as $c \to \infty$.

*Proof.* Each $G_i \sim N(0,1)$ (Theorem 7.1), so $\mathbb{E}[G_i^+] = 1/\sqrt{2\pi}$, $\mathrm{Var}(G_i^+) = 1/2 - 1/2\pi$, and $\mathrm{Corr}(G_1^+, G_2^+) = (\mathbb{E}[g(U)] - 1/2\pi)/(1/2 - 1/2\pi)$, where $g(u) = \mathbb{E}[Z_1^+ Z_2^+]$ is the positive–part covariance of a correlation–$u$ bivariate normal and $U \sim \min(\mathrm{Exp}(4c),1)$ (Theorem 7.1). By Lemma 7.2, $g'(u) = 1/4 + \arcsin u/2\pi$ and $g(0) = 1/2\pi$, so $\mathbb{E}[g(U)] - 1/2\pi = \int_0^1 g'(s)\, \mathbb{P}(U > s)\, \mathrm{d}s = 1/2\pi \int_0^1 (\pi/2 + \arcsin s)\, e^{-4cs}\, \mathrm{d}s$. Integrating by parts and using $\int_0^1 e^{-4cs} (1 - s^2)^{-1/2}\, \mathrm{d}s = \int_0^{\pi/2} e^{-4c\sin\theta}\, \mathrm{d}\theta = \pi/2\, (I_0(4c) - L_0(4c))$ (DLMF 11.5.2) gives the stated form; the limits as $c \to 0, \infty$ follow from $I_0(z) - L_0(z) = 2/\pi z + O(z^{-3})$. □

This is a feature of the limiting mixture, not a process observable: unlike the full–position correlation $(1 - e^{-4c})/(4c)$, which by Lemma 7.4 is the two species' $q$–Laplace current correlation, the positive part $G_i^+$ has no identified process–level counterpart. It is matched by Monte Carlo (§9.2).

That the dual pair $(X_1, X_2)$ captures the two species' fluctuations at all is the content of the following identity, the block–initial–condition analogue of Lemma 8.3.

## Lemma 7.4 (the dual pair computes the species cross–correlation)

Let $\eta^0$ be the block initial condition (bound pairs on $\{x \le 0\}$), and let $(X_1, X_2)$ be the different–species dual started from the bound pair. For every $q \in (0,1)$ and $s \ge 0$, the joint $q$–Laplace observable of the two species is a hitting probability of the dual pair,

$$\mathbb{E}_{\eta^0}\left[\prod_{i=1,2} \eta_{i,a}(s)\, q^{2\left(a + N_{a+1}^+(\eta_i(s))\right)}\right] \;=\; q^{2k}\, \mathbb{P}_{(a,a)}(X_1(s) \le 0,\; X_2(s) \le 0), \qquad (35)$$

with $k = 0$ for this block (a boundary at $b_0$ gives $k = 2b_0$).

*Proof.* Write $\xi^{\mathrm{bp}}$ for the dual configuration with one particle of each species at $a$. Evaluating the triangular duality function of Corollary 2.6 there leaves only the factors at the occupied site, and $N_{a-1}^-(\xi_i^{\mathrm{bp}}) = 0$, so $D^{\mathrm{tri}}(\xi^{\mathrm{bp}}, \eta) = \prod_i \eta_{i,a}\; q^{2(a + N_{a+1}^+(\eta_i))}$; the left side of (35) is thus $\mathbb{E}_{\eta^0}[D^{\mathrm{tri}}(\xi^{\mathrm{bp}}, \eta_s)]$. By the self–duality of Corollary 2.6 this equals $\mathbb{E}_{\xi^{\mathrm{bp}}}[D^{\mathrm{tri}}(\xi_s, \eta^0)]$, with $\xi_s = (X_1(s), X_2(s))$ the dual pair run from $\xi^{\mathrm{bp}}$. Since $\eta^0$ is a full block of bound pairs on $\{x \le 0\}$, $\mathbf{1}\{\xi_s \subseteq \eta^0\} = \mathbf{1}\{X_1, X_2 \le 0\}$; and on that event the exponent of each surviving factor, $x - N_{x-1}^-(\xi_{s,i}) + N_{x+1}^+(\eta_i^0)$, is independent of the particle position, because the block contributes $N_{x+1}^+(\eta_i^0) = (\text{boundary}) - x$, cancelling the $x$, and $N_{x-1}^-(\xi_{s,i}) = 0$. The weight is therefore a constant $q^{2k}$ on $\{X_1, X_2 \le 0\}$, whence $\mathbb{E}_{\xi^{\mathrm{bp}}}[D^{\mathrm{tri}}(\xi_s, \eta^0)] = q^{2k}\, \mathbb{P}_{(a,a)}(X_1, X_2 \le 0)$, which is (35).

It remains to justify the self–duality step and to identify $k$. Work first on the finite lattice $\Lambda_L = [-L, L]$ with reflecting boundaries and the truncated block $\eta^{0,L}$ (bound pairs exactly on $[-L, 0]$). The interlacing $\mathcal{L}\, D^{\mathrm{tri}} = D^{\mathrm{tri}}(\mathcal{L}^{\mathrm{dual}})^\top$ of Corollary 2.6 is a sum of bondwise identities and therefore holds for the $\Lambda_L$–generators; iterating it, $\mathcal{L}^m D^{\mathrm{tri}} = D^{\mathrm{tri}}((\mathcal{L}^{\mathrm{dual}})^\top)^m$ for every $m \ge 0$, and summing the (finite–dimensional, entrywise absolutely convergent) exponential series yields the semigroup duality $e^{s\mathcal{L}} D^{\mathrm{tri}} = D^{\mathrm{tri}} e^{s(\mathcal{L}^{\mathrm{dual}})^\top}$, i.e. the identity used above, for every $s$. For the constant: on the event $\{\xi_s \subseteq \eta^{0,L}\} = \{X_1, X_2 \leq 0\}$ each surviving factor of $D^{\mathrm{tri}}(\xi_s, \eta^{0,L})$ carries the exponent $x - N^-_{x-1}(\xi_{s,i}) + N^+_{x+1}(\eta^{0,L}_i) = x - 0 + (-x) = 0$, since the block occupies precisely the $-x$ sites $\{x+1, \dots, 0\}$ to the right of $x$; hence $D^{\mathrm{tri}}(\xi_s, \eta^{0,L}) = \mathbf{1}\{X_1(s), X_2(s) \leq 0\}$ and $k = 0$ for the block with boundary at the origin (a boundary at $b_0$ shifts each exponent to $b_0$ and gives $k = 2b_0$).

Finally let $L \to \infty$ at fixed $s$ and $a$. By the standard finite–propagation coupling, the $\Lambda_L$– and $\mathbb{Z}$–processes started from the block agree on $[-L/2, L/2]$ up to probability $O(e^{-c_0 L})$, $c_0 > 0$; the observable on the left of (35) can differ under the coupling only through particles beyond distance $L/2$ from the origin at time $s$, an event of superexponentially small probability from block initial data, and the hitting probability on the right converges by the same coupling applied to the two–particle dual. Hence (35) holds on $\mathbb{Z}$ with $k = 0$. An exact numerical evaluation of both sides for $L \leq 6$ (matrix exponentiation) confirms the identity to machine precision. □

In the regime–(A) scaling $q = 1 - c/T$, $s = T$, identity (35) expresses the two species' limiting cross–correlation through the dual pair, which Proposition 7.8 and Theorem 7.9 then compute: $\rho(c) = \lim_T \mathrm{Corr}(X_1, X_2)$.

### Lemma 7.5 (split time)

The first split time $\tau$ is $\mathrm{Exp}(\nu_{\mathrm{sp}})$, with split rate $\nu_{\mathrm{sp}} = 2q^2\varepsilon$ (recall $\varepsilon = 1 - q^2$); and $\nu_{\mathrm{sp}} T \to 4c$, $\tau/T \Rightarrow V \sim \mathrm{Exp}(4c)$, $\mathbb{P}(\tau > T) \to e^{-4c}$.

*Proof.* While bound the pair undergoes only the rate 1 and $q^4$ hops (which preserve the bound phase) and the two splits (each rate $q^2\varepsilon$, leaving it); the bound phase is a single state whose only phase–changing exit is the split, so its holding time is $\mathrm{Exp}(2q^2\varepsilon)$. The limits follow from $\varepsilon = 2c/T + O(T^{-2})$. □

### Lemma 7.6 (no re–binding)

The expected number of merges in $[\tau, T]$ is $O(c/\sqrt{T}) \to 0$.

*Proof.* After the split, $R$ is (away from a merge) a symmetric rate–$(1 + q^2)$ walk; a merge occurs only from $R = \pm 1$, at rate $\varepsilon$. The expected occupation of $\{\pm 1\}$ up to $T$ is $O(\sqrt{T})$ by Lemma 4.1 (with re–binding episodes handled by the excursion bound in the proof of Lemma 4.2, which contributes only the factor $1 + \nu_{\mathrm{sp}} T = O(1 + c)$), so the expected number of merges is $\leq \varepsilon \cdot O(\sqrt{T}) = O(c/\sqrt{T})$ for fixed $c$. □

It is convenient to pass to the sum and difference $S := X_1 + X_2$, $R := X_2 - X_1$, so that $X_i = (S \mp R)/2$ and the interaction (confined to the adjacent set $\{|R| = 1\}$ and the origin) acts only on $R$.

### Lemma 7.7 (exact symmetry identities and adjacent occupation)

Species–interchange symmetry maps $R \mapsto -R$, $S \mapsto S$ and fixes the law (the initial bound pair is symmetric); consequently

$$\mathbb{E}[R(t)] = 0, \qquad \mathrm{Cov}(S(t), R(t)) = 0 \qquad (\forall t \geq 0). \qquad (36)$$

Moreover, on the event $A_T := \{\tau \leq T,\ \text{no re–binding}\}$ — on which, after the split, $R$ performs the no–merge walk — the adjacent–set occupation $\Lambda_T := \int_\tau^T \mathbf{1}\{|R(t)| = 1\}\, \mathrm{d}t$ has conditional expectation $\mathbb{E}[\Lambda_T \mid A_T] = O(\sqrt{T})$; in particular $\Lambda_T = O_{\mathbb{P}}(\sqrt{T})$. (Throughout, $Y_T = O_{\mathbb{P}}(a_T)$ means the family $Y_T/a_T$ is tight — for every $\varepsilon > 0$ there is $M$ with $\sup_T \mathbb{P}(|Y_T| > M a_T) \leq \varepsilon$ — and $Y_T = o_{\mathbb{P}}(a_T)$ means $Y_T/a_T \to 0$ in probability.)

*Proof*. The identities (36) follow from $X_i = (S \mp R)/2$ and the invariance of the law of the process under $R \mapsto -R$. After the split and with no merge, $R$ is a continuous–time walk on $\mathbb{Z}\backslash\{0\}$ of total rate $\leq 2(1+q^2)$, symmetric on $\{|R| \geq 2\}$; by Lemma 4.1 its time–$t$ marginal is $\leq C/\sqrt{1+t}$ uniformly, so the expected occupation of $\{\pm 1\}$ up to $T$ under this no–merge law — which is $\mathbb{E}[\Lambda_T \mid A_T]$ — is $\int_0^T C/\sqrt{1+t}\, \mathrm{d}t = O(\sqrt{T})$. □

## Proposition 7.8 (two–phase functional CLT)

Conditionally on $\tau/T \to u \in [0,1]$ and on $A_T$,

$$\frac{1}{\sqrt{2T}}(X_1(T), X_2(T)) \Rightarrow \left(\sqrt{u}\,\xi_0 + \sqrt{1-u}\,\xi_1,\ \sqrt{u}\,\xi_0 + \sqrt{1-u}\,\xi_2\right),$$

with $\xi_0, \xi_1, \xi_2$ i.i.d. $N(0,1)$, a standard bivariate normal of correlation $u$. Since $U := \min(\tau/T, 1) \Rightarrow \min(\mathrm{Exp}(4c), 1)$ (Lemma 7.5), the limiting law of $(X_1, X_2)/\sqrt{2T}$ is that of $\left(\sqrt{U}\,\xi_0 + \sqrt{1-U}\,\xi_1,\ \sqrt{U}\,\xi_0 + \sqrt{1-U}\,\xi_2\right)$ with $U \sim \min(\mathrm{Exp}(4c), 1)$ drawn independently of $\xi_0, \xi_1, \xi_2$.

*Proof*. (Conditioning on $A_T$ is innocuous: $\mathbb{P}(A_T^c \mid \tau \leq T) \to 0$ by Lemma 7.6, so the conditioned and unconditioned post–split laws agree to vanishing total variation; we may therefore compute with the no–merge rates.)

<u>Pre–split.</u> On $[0, \tau]$ the common pair position $D$ is a walk of right rate 1 and left rate $q^4$, hence variance rate $1 + q^4 \to 2$ and drift $1 - q^4 = O(c/T)$. By Donsker's invariance principle with $\tau \approx uT$, $D(\tau)/\sqrt{2T} \Rightarrow \sqrt{u}\,\xi_0$ (the drift contributes $(1-q^4)\tau/\sqrt{2T} = O(c\sqrt{u}/\sqrt{T}) \to 0$). Since $S(\tau) = 2D(\tau) + O(1)$ and $R(\tau) = O(1)$, the split deposits $S(\tau)/\sqrt{2T} \Rightarrow 2\sqrt{u}\,\xi_0$ and $R(\tau)/\sqrt{2T} \to 0$.

<u>Post–split.</u> Write $\bar{S}(s) = S(\tau + s) - S(\tau)$, $\bar{R}(s) = R(\tau + s) - R(\tau)$; each is a compensated jump process. Reading the rates of §5, the instantaneous drifts vanish for $R$ on $\{|R| \geq 2\}$ and equal $2\varepsilon$ for $S$, while on $\{|R| = 1\}$ they are $O(\varepsilon)$ for $R$ and $\varepsilon$ for $S$ (the deficit $2\varepsilon \to \varepsilon$ is the excluded merge); hence $\mathbb{E}[\bar{R}(T-\tau)]$ is supported on $\{|R| = 1\}$ with magnitude $\leq \varepsilon\,\mathbb{E}[\Lambda_T] = O(c/\sqrt{T})$, and $\mathbb{E}[\bar{S}(T-\tau)] = 2\varepsilon(T-\tau) - \varepsilon\,\mathbb{E}[\Lambda_T] = O(c)$, both $o(\sqrt{T})$. The predictable quadratic–variation rates are, on $\{|R| \geq 2\}$,

$$a_S = a_R = 2(1+q^2), \qquad a_{SR} = 0,$$

and on $\{|R| = 1\}$ the bounded values $a_S = 1 + q^2$, $a_R = 1 + 5q^2$, $a_{SR} = \varepsilon$ (the swap $R = \pm 1 \mapsto \mp 1$ contributes $\Delta R = \mp 2$). Therefore

$$\langle \bar{S} \rangle = 2(1 + q^2)(T - \tau) - (1 + q^2)\Lambda_T, \ \ \langle \bar{R} \rangle = 2(1 + q^2)(T - \tau) + (3q^2 - 1)\Lambda_T, \ \ \langle \bar{S}, \bar{R} \rangle = \varepsilon\, \Lambda_T,$$

and by Lemma 7.7 every $\Lambda_T$–term is $O_{\mathbb{P}}(\sqrt{T}) = o_{\mathbb{P}}(T)$. With $q^2 \to 1$ and $(T - \tau)/T \to 1 - u$,

$$1/2T\, \langle \bar{S} \rangle \to 2(1 - u), \quad 1/2T\, \langle \bar{R} \rangle \to 2(1 - u), \quad 1/2T\, \langle \bar{S}, \bar{R} \rangle \to 0.$$

The jumps are bounded by 2, so the martingale functional CLT [76], Thm. 7.1.4 gives $(\bar{S}, \bar{R})(T - \tau)/\sqrt{2T} \Rightarrow (\sqrt{2(1 - u)}\, \eta_S, \sqrt{2(1 - u)}\, \eta_R)$ with $\eta_S, \eta_R$ i.i.d. $N(0,1)$, independent of $\mathcal{F}_\tau$ (hence of $\xi_0$) by the strong Markov property at $\tau$.

Recombination. Adding the two phases, $S(T)/\sqrt{2T} \Rightarrow 2\sqrt{u}\, \xi_0 + \sqrt{2(1 - u)}\, \eta_S$ and $R(T)/\sqrt{2T} \Rightarrow \sqrt{2(1 - u)}\, \eta_R$. Setting $\xi_1 = (\eta_S - \eta_R)/\sqrt{2}$, $\xi_2 = (\eta_S + \eta_R)/\sqrt{2}$ (an orthogonal change of variables, so $\xi_0, \xi_1, \xi_2$ are i.i.d. $N(0,1)$) and using $X_i = (S \mp R)/2$ gives the claim. Averaging over $U = \min(\tau/T, 1) \Rightarrow \min(\mathrm{Exp}(4c),1)$ (Lemma 7.5) gives the unconditional limit. □

### Theorem 7.9 (closed form for the crossover correlation)

As $T \to \infty$, $\mathrm{Corr}(X_1, X_2) \to (1 - e^{-4c})/(4c)$, decreasing from 1 as $c \to 0$ to a $1/c$ tail as $c \to \infty$.

*Proof.* By the exact identity $\mathrm{Cov}(S, R) = 0$, $\mathrm{Corr}(X_1, X_2) = (\mathrm{Var}S - \mathrm{Var}R)/(\mathrm{Var}S + \mathrm{Var}R)$. The per–state increment second moments of $S, R$ satisfy $a_r + c_r = 4(1 + q^2)$ for every internal state $r$ (a direct check of the rates), so $\mathrm{Var}S + \mathrm{Var}R = 4(1 + q^2) \sum_r \tau_r = 4(1 + q^2)T$ up to the drift–compensator variances, which are $O(1)$ (the compensators are $O(\varepsilon)$–Lipschitz functionals of occupations) and are absorbed in the error terms below, with $\tau_r$ the occupation of state $r$; and $a_r - c_r$ is supported on $\{0, \pm 1\}$ with $a_0 - c_0 = 4(1 + q^4)$, $a_{\pm 1} - c_{\pm 1} = -4q^2$, so $\mathrm{Var}S - \mathrm{Var}R = 4(1 + q^4)\tau_0 - 4q^2(\tau_{+1} + \tau_{-1})$. In the scaling limit the bound–state occupation $\tau_0$ is the first–split time (re–binding negligible, Lemma 7.6), $\mathbb{E}[\tau_0]/T \to (1 - e^{-4c})/(4c)$, and the $\{|r| = 1\}$ occupations are $o(T)$; hence $\mathrm{Corr} \to (1 - e^{-4c})/(4c)$. □

## 8 The Tracy–Widom regime

Now fix $q \in (0,1)$ and study the integrated currents $N_1(s), N_2(s)$ across the origin from the step initial condition at a single time $s$, with standardised currents $\mathcal{O}^i = (N_i - \mathbb{E}N_i)/\sigma_i$, $\sigma_i \asymp s^{1/3}$. The marginal statements hold for every $n$ by Proposition 3.1, via the time–changed generator $\tilde{\mathcal{L}}^{(n)} = \beta_n^{-1}\mathcal{L}^{(n)}$, under which the single–species rates are the $n$–free $q^{\mp 1}$. The joint statement (Theorem 8.4) uses the duality and is proved at $n = \infty$; it extends verbatim to $n = 2{,}3$, where the duality is known [48], and to every $n$ modulo the duality conjecture of [48], Rmk. 4.

## 8.1 Marginals

*Theorem 8.1 (Tracy–Widom marginals)*

For every $n \in \mathbb{N} \cup \{\infty\}$ and each species, $\mathcal{O}_s^i \Rightarrow F_2$, the GUE Tracy–Widom law.

*Proof.* By Proposition 3.1(a), each species' current is an autonomous single–species ASEP — with rates $r_R = 1, r_L = q^2$ at $n = \infty$ and $q^{\mp 1}\beta_n$ at finite $n$, the factor $\beta_n$ a pure time change removed by the standardisation. The step block (state 3 on $\{x \le 0\}$, empty on $\{x > 0\}$) projects, per species, to the step initial condition, and step–ASEP currents are Tracy–Widom by [57], Thm. 1.4. □

*Remark 8.2*

Theorem 8.1 proves the Tracy–Widom conjecture of [48, Sec. 3.1.1] (stated there for $n = 2,3$; here for every $n$), up to the standard equivalence between current and tagged–position asymptotics.

## 8.2 Triangular duality and $q$–Laplace decoupling

For the cross sector the dual has one walker of each species; let $\mathfrak{r} = X_2 - X_1$ be the relative coordinate, which is driftless (the two species share the drift), finite–range, with a bounded perturbation at the origin from the binding. The triangular duality of Corollary 2.6 pairs finite dual configurations with the step sector, on which it takes a strikingly simple form.

*Lemma 8.3 (step–sector duality identities)*

Let $\eta^0$ be the step (state 3 on $x \le 0$, empty on $x > 0$), let $X(\cdot)$ denote the free asymmetric walk (rates 1 right, $q^2$ left), and let $(X_1, X_2)$ denote the two–particle dual — one particle of each species, each marginally a free walk by Proposition 3.1, jointly carrying the binding interaction on the contact set $\{|X_1 - X_2| \le 1\}$. Then for all $a, b \in \mathbb{Z}$ and $s \ge 0$,

$$\mathbb{E}_{\eta^0}\Big[\eta_{i,a}(s)\, q^{2\left(a+N_{a+1}^{+}(\eta_i(s))\right)}\Big] = \mathbb{P}_a(X(s) \le 0), \tag{37}$$

$$\mathbb{E}_{\eta^0}\Big[\eta_{1,a}(s)\, \eta_{2,b}(s)\, q^{2\left(a+N_{a+1}^{+}(\eta_1(s))\right)+2\left(b+N_{b+1}^{+}(\eta_2(s))\right)}\Big] = \mathbb{P}_{(a,b)}(X_1(s) \le 0,\ X_2(s) \le 0), \tag{38}$$

and $\mathbb{P}_{(a,b)}(X_1(s) \le 0) = \mathbb{P}_a(X(s) \le 0)$ exactly.

*Proof.* Apply Corollary 2.6 with $\xi$ a single species–$i$ dual particle at $a$ (respectively the pair at $(a,b)$). With one particle per species, $N^- \equiv 0$; and on the step, $N_{z+1}^{+}(\eta_i^0) = -z$ for $z \le 0$, so every weight in $D^{\mathrm{tri}}(\xi_s, \eta^0)$ equals $q^{2(X_i - X_i)} = 1$ and $D^{\mathrm{tri}}(\xi_s, \eta^0) = \mathbf{1}\{X_i(s) \le 0\ \forall i\}$. On a finite lattice the displays are then exactly the duality relation; the extension to $\mathbb{Z}$ is the standard finite–propagation–speed approximation: couple the infinite system to its restriction to $[-K, K]$; for fixed $s$ the two agree on $[-K/2, K/2]$ up to probability $O(e^{-c_0 K})$, $c_0 > 0$, and (since the initial condition is the step) the event that any particle lies beyond $K/2$ at time $s$ — the only way the right–tail counts $N^+$ in the observables can disagree — has super–exponentially small probability. We have additionally verified (37)–(38) exactly, by matrix exponentiation of the full forward dynamics on 8 sites against the one– and two–particle dual sectors. The marginal consistency is Proposition 3.1: the species–1 coordinate of the dual pair is an autonomous free walk. □

The identities (37)–(38) show precisely what the two–particle dual computes: the currents enter through the $q$–deformed weights $q^{2N^+}$, not linearly. Summing the telescope

$$q^{2N^+_{a+1}(\eta_i)} - q^{2N^+_a(\eta_i)} = (1-q^2)\,\eta_{i,a}\,q^{2N^+_{a+1}(\eta_i)} \qquad (39)$$

over half–lines converts them into exact joint $q$–Laplace formulas, and the decoupling theorem of this regime is the following contact representation.

*Theorem 8.4 (q–Laplace decoupling: exact contact representation)*

At $n=\infty$, from the step initial condition, for all $s \ge 0$,

$$\mathbb{E}\left[q^{2N_i(s)}\right] = 1 - (1-q^2)\sum_{a\ge 1} q^{-2a}\ \mathbb{P}_a(X(s)\le 0), \qquad (40)$$

$$\mathrm{Cov}\left(q^{2N_1(s)},\ q^{2N_2(s)}\right) = (1-q^2)^2 \sum_{a,b\ge 1} q^{-2(a+b)}\ C_s(a,b), \qquad (41)$$

where $C_s(a,b) := \mathbb{P}_{(a,b)}(X_1(s)\le 0, X_2(s)\le 0) - \mathbb{P}_a(X(s)\le 0)\mathbb{P}_b(X(s)\le 0)$ is the connected pair kernel and all sums converge absolutely. The kernel is generated entirely by contact: $|C_s(a,b)| \le \mathbb{P}_{(a,b)}(\tau_c \le s)$, with $\tau_c$ the first meeting time of the two dual walkers, and $C_s \equiv 0$ if the binding interaction is removed. In particular the two currents decouple in the $q$–Laplace sense: $|\mathrm{Cov}(q^{2N_1}, q^{2N_2})(s)| \le \mathbb{E}\left[q^{2N_1(s)}\right] \to 0$.

*Proof.* (39) is $N^+_a = N^+_{a+1} + \eta_{i,a}$; summed over $a \ge m$ it telescopes to $1 - q^{2N^+_m(\eta_i)}$ (the configuration is right–finite, so $N^+_a \to 0$), and $N^+_1(\eta_i) = N_i$. Taking expectations at $m=1$ and inserting (37) weighted by $q^{-2a}$ gives (40); the double telescope with (38) gives

$$\mathbb{E}\left[\frac{(1-q^{2N_1})(1-q^{2N_2})}{(1-q^2)^2}(s)\right] = \sum_{a\ge 1}\sum_{b\ge 1} q^{-2(a+b)}\ \mathbb{P}_{(a,b)}(X_1(s)\le 0,\ X_2(s)\le 0),$$

and expanding the product and subtracting the marginal identities yields (41). Absolute convergence: to reach $\{\le 0\}$ from $a$ the walk makes at least $a$ jumps, so $\mathbb{P}_a(X(s)\le 0) \le \mathbb{P}(\mathrm{Poi}((1+q^2)s) \ge a)$ decays super–exponentially in $a$ for fixed $s$, beating $q^{-2a}$; this justifies Fubini throughout. Contact generation: realise the pair $(X_1, X_2)$ and an independent pair $(\tilde{X}_1, \tilde{X}_2)$ on one probability space, with identical trajectories until the first time $\tau_c$ at which the walkers come within distance 1 — possible since off the contact set the pair dynamics consists of the same independent free jumps. Then $C_s(a,b) = \mathbb{E}[\mathbf{1}\{X_1(s)\le 0, X_2(s)\le 0\} - \mathbf{1}\{\tilde{X}_1(s)\le 0, \tilde{X}_2(s)\le 0\}]$, which vanishes on $\{\tau_c > s\}$; hence $|C_s(a,b)| \le \mathbb{P}(\tau_c \le s)$, and $C_s \equiv 0$ for independent walkers. Finally, for $U, V \in [0,1]$ one has $|\mathrm{Cov}(U,V)| \le \min(\mathbb{E}U, \mathbb{E}V)$, and $\mathbb{E}[q^{2N_1(s)}] \to 0$ since $N_1(s) \to \infty$ in probability (the hydrodynamic limit of the autonomous ASEP marginal from step has strictly positive current) and $q^{2N_1} \le 1$. □

*Conjecture 8.5 (linear covariance)*

At every $n$, from the step initial condition, $\mathrm{Cov}(N_1, N_2)(s) \sim c(q)\sqrt{s}$ as $s \to \infty$ with $c(q) \in (0, \infty)$; consequently $\mathrm{Corr}(N_1, N_2) \asymp s^{-1/6} \to 0$ and the two Tracy–Widom currents are asymptotically uncorrelated. At $q = 1/2$, Monte Carlo gives $c = 0.099 \pm 0.003$, with the $\sqrt{s}$ scaling holding over $s \in [50, 3.3 \times 10^4]$ and the correlation decaying at the $s^{-1/6}$ rate (§9.2).

The linear covariance is not a finite–mass dual observable: by (37)–(38) the duality reaches the currents only through the weights $q^{2N^+}$, and extracting $\mathbb{E}[N_1 N_2]$ requires the full tower of joint $q$–moments (same–species dual particles in unbounded number) — the same difficulty as the joint law itself, left open in §9.3. This, and not a missing estimate, is why the covariance asymptotics are stated as a conjecture.

The natural candidate for the conjectured coefficient $c(q)$ is the contact combination $(1 + q^4)\tau_0 - q^2(\tau_{+1} + \tau_{-1})$, $\tau_r = \int_0^s \mathbb{P}\,(\mathfrak{r}(t) = r)\,\mathrm{d}t$, since (exactly as in the proof of Theorem 7.9) the per–state increment second moments of $\mathfrak{s} = X_1 + X_2$ and $\mathfrak{r}$ coincide off contact ($a_r = c_r$ for $|r| \ge 2$) while at contact $a_0 - c_0 = 4(1 + q^4)$, $a_{\pm 1} - c_{\pm 1} = -4q^2$.

*Proposition 8.6 (contact occupations of the relative walk)*

The relative walk $\mathfrak{r}$ is reversible with respect to the measure $m(0) = q^{-2}$, $m(r) = 1$ $(r \ne 0)$: detailed balance across the bond $\{0,1\}$ reads $m(0) \cdot q^2\varepsilon = m(1) \cdot \varepsilon$ (split out of the bound state against merge into it), and is immediate at the swap and at all bulk bonds. Consequently

$$\frac{\tau_0(s)}{\tau_{\pm 1}(s)} \;\to\; q^{-2} \qquad (s \to \infty),$$

the bound state being overweighted by exactly the stickiness ratio $q^{-2}$, and

$$(1 + q^4)\,\tau_0 - q^2(\tau_{+1} + \tau_{-1}) \;\sim\; \frac{1 - q^4}{q^2}\sqrt{\frac{s}{\pi(1 + q^2)}}.$$

*Proof.* Reversibility: the bulk rates ($1 + q^2$ each way) and the swap ($q^2$ each way between $\pm 1$) are symmetric, so $m$ is constant off the origin; the only nontrivial bond is $0 \leftrightarrow \pm 1$, with rate $q^2\varepsilon$ out (split) and $\varepsilon$ in (merge), giving $m(0) = q^{-2} m(\pm 1)$. The walk is recurrent (off a finite set it is the symmetric nearest–neighbour walk) and satisfies the hypotheses of Lemma 4.6 with $a = 1 + q^2$, the swap being part of the finite defect, so $\tau_r(s) \sim m(r)\sqrt{s/(\pi(1 + q^2))}$, and the two displays follow. Numerically (exact integration of the master equation at $q = 1/2$): $p_t(0,0)/p_t(0,1) = 4.002$ at $t = 3000$ against $q^{-2} = 4$, and the combination divided by $\sqrt{s}$ is 1.875 at $s = 3000$ against the predicted limit 1.892, the deficit decaying as $s^{-1/2}$. □

*Remark 8.7 (why there is no ballistic term)*

The step block is deterministic and species–$i$ particles enter $\{x > 0\}$ only across the bond $(0,1)$, so $\delta N_i = \delta J_i$ with $J_i$ the single–bond integrated current; this localization gives the transport reading. Decomposing $J_i = M_i + A_i$ (martingale plus compensator), the martingale cross–bracket is exactly the contact carré du champ $\langle M_1, M_2\rangle(s) = \int_0^s V_0\,(\eta_u)\,\mathrm{d}u$ with $V_0$ of (14) — only bound–pair hops and swaps move both species across one bond. The drift cross–terms $\mathrm{Cov}(M_i, A_j), \mathrm{Cov}(A_1, A_2)$ have a leading part equal to the equilibrium Green–Kubo cross–mobility $\sigma_{12} \cdot s$, which vanishes by Proposition 3.2. So $\mathrm{Cov}(N_1, N_2)$ carries no $O(s)$ ballistic part, and the surviving $\sqrt{s}$ is the transient contact contribution evaluated above by duality. The forward decomposition thus explains the absence of an $O(s)$ term but does not by itself yield the constant: $M_i$ and $A_i$ each have $\Theta(s)$ variance and their cancellation to the $O(\sqrt{s})$ remainder is a transport–coefficient near cancellation, in contrast to the exact, term–by–term off–contact identity $a_r = c_r$ supplied by the duality.

*Remark 8.8*

Theorem 8.4 is the strongest joint statement we prove unconditionally: an exact identity, valid for all $s$, in which the inter–species coupling enters only through the contact kernel $C_s$ — the $q$–Laplace analogue of uncorrelatedness. For the linear currents, Conjecture 8.5 is supported by the exact $\sqrt{s}$ scaling in the simulations and by the occupation analysis of Proposition 8.6; a proof would require joint control of the full $q$–moment tower, and even then says nothing about higher joint cumulants.

# 9 Phase picture, numerical evidence, and discussion

This concluding section, unlike §§6–8, is not rigorous: it places the results in a single phase picture (§9.1), corroborates them by simulation (§9.2), and discusses their scope and the open problems (§9.3).

## 9.1 The asymmetry–exponent phase picture

The two regimes are endpoints of a one–parameter family of scalings, and the following observations — established at the level of the explicit rate algebra and extensive simulation, but not with the rigour of §§6–8 — place them in a single picture. Tune

$$q = 1 - c\,T^{-\alpha}, \qquad \alpha \geq 0,\ c > 0, \qquad (42)$$

and observe the integrated currents at time $T$; the splits per particle are $\sim T(1-q^2) \sim 2cT^{1-\alpha}$, governing the limiting cross–correlation.

- $\alpha < 1$: $\to \infty$ splits per particle, the species mix, $\mathrm{Corr}(N_1, N_2) \to 0$; the marginals are Tracy–Widom (the asymmetry is still macroscopic), and $\alpha = 0$ is the fixed–$q$ regime of §8.

- $\alpha = 1$ (critical): $O(c)$ splits, $\mathrm{Corr} \to \rho(c) \in (0,1)$, the crossover of Theorem 7.9; $\gamma \sim c/T$ vanishes, the KPZ nonlinearity switches off, and the limit is the Edwards–Wilkinson pair of §6.

- $\alpha > 1$: $\to 0$ splits, bound pairs survive, $\mathrm{Corr} \to 1$ continuously — the locked limit of Remark 2.3.

A single axis thus interpolates from decoupled Tracy–Widom ($\alpha = 0$) through the EW crossover ($\alpha = 1$) to locking ($\alpha \to \infty$); the hydrodynamics decouple for every $\alpha, n$ (Proposition 3.1), only the fluctuation–level coupling depending on the scaling. The crossover correlation $\rho(c)$ is not a transport coefficient: $C_{12} \equiv 0$ and (numerically) $\sigma_{12} \approx 0$ (measured $-0.024 \pm 0.022$ at $c = 1$, against the front value $\rho(1) \approx 0.27$), so what remains is a transient, shared–source correlation — the species released by a common split carrying correlated currents, $\rho(c)$ the decaying memory of the perfectly correlated block initial condition. To our knowledge this combination (Edwards–Wilkinson class, diagonal compressibility, vanishing cross–mobility, nonzero initial–condition–driven cross–correlation with a $\rho(c) \sim 1/c$ crossover) matches no model in the multi–species fluctuating hydrodynamics or macroscopic fluctuation theory literature.

We record three further numerical observations, consistent across $n$ (and into the negative–$n$ window of Remark 2.3):

- Crossover time. At fixed $q$ near 1 the Tracy–Widom statistics emerge only past $t_{\mathrm{cross}} \sim (1-q)^{-3/2}$; observed earlier the marginal looks Gaussian (the $\alpha \geq 1$ behaviour at finite time). At $q = 1/2$ the law is clean by $T \sim 10^3$ (skewness $\to -0.21$, the $F_2$ value).

- Subleading variance. $\mathrm{Var}(N_i) = A\, s^{2/3} + B\, s^{1/2} + \cdots$, a relative–mode (Popkov–Schütz–type) correction; the standardised remainder $\sigma_{\mathrm{extra}}^2 \asymp s^{-1/6}$ decays at the same rate as the cross correlation ($\sigma_{\mathrm{extra}} \approx 0.39$ vs $0.36$ at $n = 2$ vs $13$, $T = 3.3 \times 10^4$).

- Multipoint uncorrelatedness. The cross–species height functions are uncorrelated at all spatial separations ($n = 2$, $q = 1/2$, $T = 3.3 \times 10^4$), so the decoupling extends from the integrated current to the spatial process, and it is $n$–independent at that level.

## 9.2 Numerical evidence

We summarise the Monte Carlo simulations that corroborate the theorems and the phase picture. Lattice simulations use the exact $n = \infty$ rate table (3) from a block initial condition of bound pairs; dual computations use the two– and few–particle generators directly. All runs store only currents and scalars (not lattice states), with $5 \times 10^4$ trials unless noted.

*The Edwards–Wilkinson regime.*

The single–species current standard deviation fits the EW law $\mathrm{std}(N_1) \sim T^{1/4}$ across $T \in [100, 10^4]$ (local slopes 0.22–0.28, fit exponent 0.244), excluding the $T^{1/2}$ that a bound–pair third field would produce, consistent with the autonomous single–species drift of Proposition 6.18. The initial–condition crossover matches the closed forms of Theorem 7.9 to sampling error, with the predicted $1/c$ tail (Figure 1); the simpler full–position observable $(1 - e^{-4c})/(4c)$ matches equally well. Direct measurement gives $\mathrm{Corr}(J_1, J_2) = -0.02 \pm 0.02$ (vanishing cross–mobility) and $C_{12} = 0$ within error, confirming that $\rho(c)$ is neither a static nor a dynamic transport coefficient.

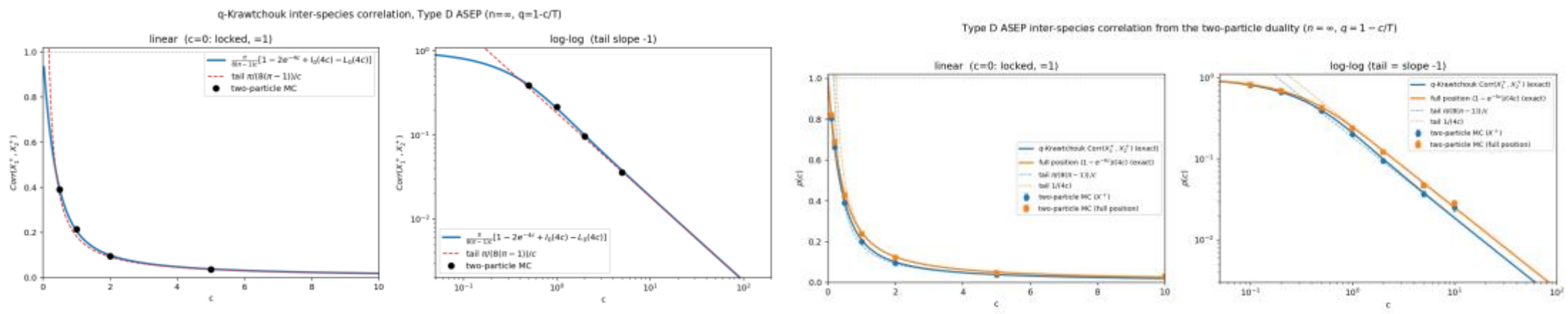

*The initial–condition crossover. Left: the one–sided correlation $Corr(X_1^+, X_2^+)$ (Proposition 7.3, Bessel–Struve closed form, curve) against two–particle Monte Carlo (dots), with the $\pi/8(\pi-1)\, c^{-1}$ tail. Right: the full–position correlation $Corr(X_1, X_2) \to (1 - e^{-4c})/(4c)$ (Theorem 7.9) against Monte Carlo. Both run from perfectly correlated data ($Corr \to 1$ as $c \to 0$) to the decoupled $1/c$ tail.*

*The Tracy–Widom regime.*

The covariance scales as $\mathrm{Cov}(N_1, N_2) \sim T^{1/2}$ (Conjecture 8.5), confirmed by exact augmented–matrix occupation–time computation ($\tau_0 \sim T^{0.505}$, $\mathrm{Cov} \sim T^{0.504}$ at $q = 1/2$) and by forward simulation (Figure 2, left). A dedicated forward run ($4.5 \times 10^4$ trials, $T \leq 800$, $q = 1/2$) pins the prefactor: $\mathrm{Cov}(N_1, N_2) = (0.099 \pm 0.003)\sqrt{T}$, in agreement with the $n = \infty$ data of Figure 2 ($\approx 0.105\sqrt{T}$ over $T \in [10^2, 10^4]$); the same run reproduces the Borodin–Corwin–Sasamoto variance $2^{-8/3}(\gamma T)^{2/3}\mathrm{Var}(\chi_2)$, $\gamma = 1 - q^2$, to within 10%, and exact integration of the relative–walk master equation gives $\tau_0/\tau_1 \to q^{-2}$ as in Proposition 8.6. The measured prefactor lies well below the contact–occupation combination of Proposition 8.6 ($\approx 1.89$ at $q = 1/2$), confirming that $\mathrm{Cov}(N_1, N_2)$ is not given by that combination but carries a nontrivial front factor. The marginals converge to $F_2$ (Figure 2, middle: CDF/histogram; skewness $\to -0.21$, excess kurtosis $\to 0.09$, the GUE Tracy–Widom values), and the limit is $n$–independent (Figure 2, right: $n = 2$ and $n = 13$ collapse onto $F_2$). The standardised correlation accordingly decays, $\mathrm{Corr}(N_1, N_2) \sim s^{-1/6} \to 0$, the asymptotic decoupling of Conjecture 8.5.

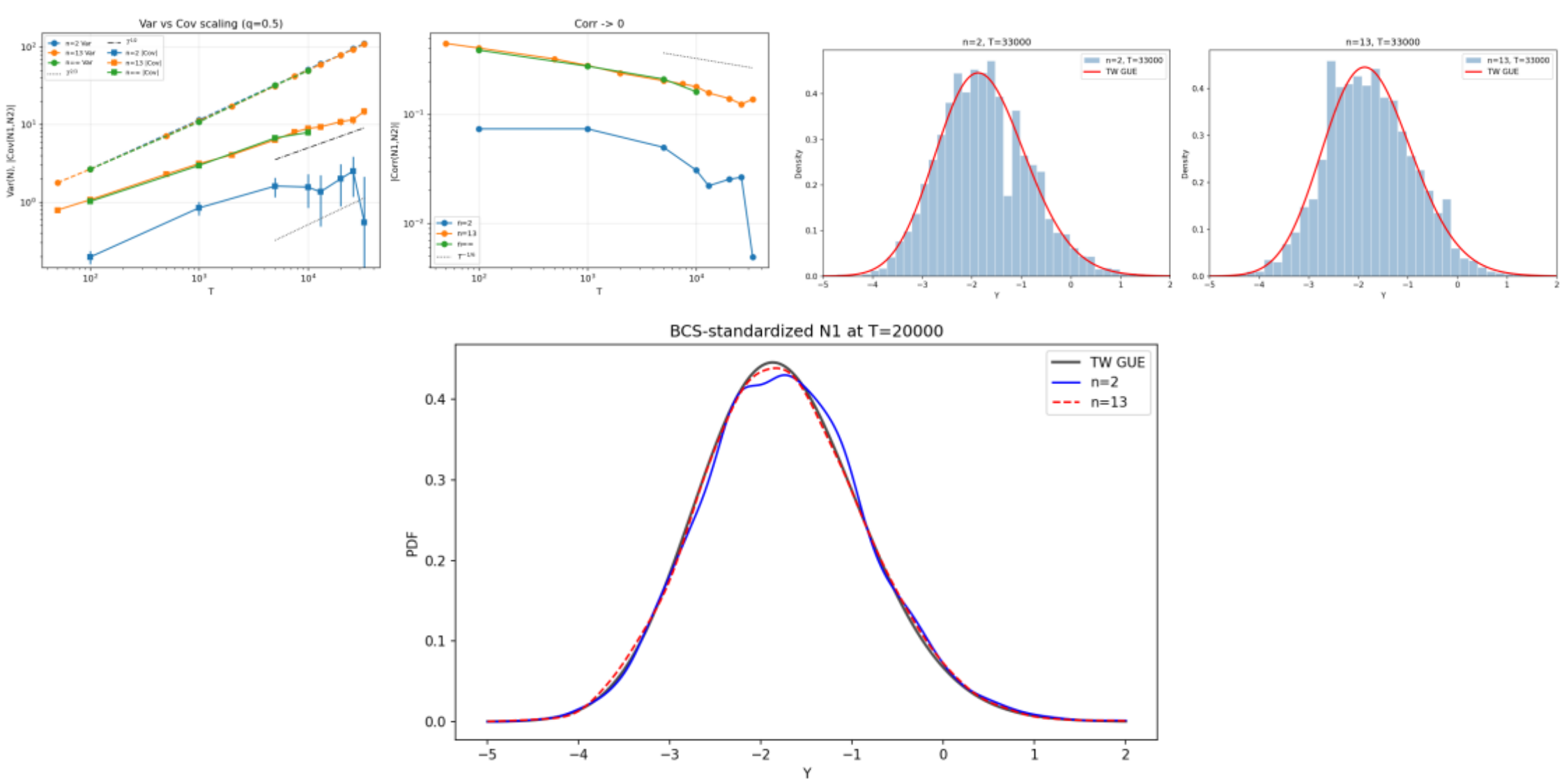

*The Tracy–Widom regime. Top left: $Cov(N_1, N_2) \sim T^{1/2}$, the uncorrelatedness of Conjecture 8.5. Top right: the marginal current against the GUE Tracy–Widom law $F_2$ at $T = 3.3 \times 10^4$. Bottom: $n$–independence — the standardised marginals at $n = 2$ and $n = 13$ collapse onto $F_2$.*

## 9.3 Discussion

The type D bound pair is a slow, near–conserved internal mode, yet in both scaling regimes the species decouple in the limit, for one structural reason (Remark 3.3): the asymmetry lives only in the marginals (Proposition 3.1), and the coupling is invisible to both equilibrium transport coefficients (Proposition 3.2). What is left of the coupling is a transient, initial–condition–driven object: in regime (A) the entire crossover $\rho(c)$ (Theorem 7.9); in regime (B) the contact kernel of the exact $q$–Laplace representation (Theorem 8.4) and, conjecturally, the sub–leading $s^{-1/6}$ correlation that decays to leave the two currents uncorrelated (Conjecture 8.5). The two regimes are governed by one object — the contact occupations $(1 + q^4)\tau_0 - q^2\tau_{\pm 1}$ of the relative walk — in two windows: at $\alpha = 1$ the split rate is $O(\varepsilon) \to 0$, $\tau_0 = \Theta(T)$, the crossover $O(1)$; at $\alpha = 0$ the split rate is $\Theta(1)$, $\tau_0 = \Theta(\sqrt{T})$, the correlation sub–leading.

We emphasise the scope: both results are statements of decoupling. In regime (A) the Gaussian limiting fields are uncoupled in drift and noise; in regime (B) the two currents decouple exactly at the level of the joint $q$–Laplace observable, and — conjecturally, with exact numerical scaling — are asymptotically uncorrelated. We do not address the finer question of whether the regime (B) limit is a product (joint) law; the higher–order joint structure of the two currents, the linear covariance asymptotics of Conjecture 8.5, and its constant $c(q)$ are facets of one open problem: the joint $q$–moment tower beyond two dual particles.

The Edwards–Wilkinson decoupling holds at arbitrary densities $(\rho_1, \rho_2)$, not only on the symmetric line (Theorem 6.5): every ingredient is density–free — the current decoupling (Proposition 3.1), the cross–noise cancellation $\mathbb{E}_\nu[V_x] = 0$ (Proposition 3.2), the orthogonality $V_x \perp$ density (Lemma 6.7), and the kernel bound (Theorem 5.2), while the diagonal noise is per species — so the two fields decouple into linear stochastic heat equations with species–dependent noise intensities $\sqrt{2\chi_i D}$, $\chi_i = \rho_i(1-\rho_i)$, and the bound–pair mode never produces a third hydrodynamic field (Corollary 6.16, Remark 6.17). The one genuinely open direction is the non–stationary EW analysis from block initial data, for which $\rho(c)$ is the exact limiting cross–correlation and Proposition 7.8 the natural tool.

*Status.*

The $n = \infty$ self–duality is Theorem 2.5(ii) (two–site interlacing verified by computer algebra, extended to all $L$ by the induction of [48]). The computer–algebra and numerical verifications cited in Theorem 2.5, Propositions 3.1 and 2.8, Lemmas 6.7, 6.12, 8.3 and 7.4, and §9.2 are reproduced by scripts. In regime (A): Propositions 3.2, 6.18, Corollary 6.16, Lemmas 6.7, 6.8, 4.1–4.6, 5.1, 7.5, 7.6, Theorems 5.2, 7.9 and Proposition 7.8 are complete and self–contained; Lemmas 6.9 and 6.12 are complete and self–contained (an exact telescoping bound at compensated fugacity, and an exact identity plus the projection inequality, respectively); Proposition 6.14 rests on the orthogonality of Proposition 2.8 (proved via [63]) and is complete modulo the infinite–volume step recorded in Remark 2.9; Theorem 6.5 is complete modulo the classical single–species WASEP fluctuation result (used for the drift via Proposition 3.1 and the diagonal noise) and the classical tightness/uniqueness [71]. In regime (B): Theorem 8.1 (marginals, every $n$, via Proposition 3.1) is complete; Theorem 8.4 ($q$–Laplace decoupling) is complete at $n = \infty$ — the triangular reduction is Lemma 8.3 — extends to $n = 2,3$ via [48], and for other finite $n$ rests on the duality conjecture of [48], Rmk. 4; the linear–current statement is Conjecture 8.5, its constant open. Section 9.1 is numerical/exploratory and frames the rigorous results within the $(\alpha, q, n)$ picture.

## Bibliography

1. H. Spohn, “Nonlinear fluctuating hydrodynamics for anharmonic chains,” *J. Stat. Phys.* ***154*** *(2014), 1191–1227; arXiv:1305.6412*, 2014.
2. P. L. Ferrari, T. Sasamoto, and H. Spohn, “Coupled Kardar–Parisi–Zhang equations in one dimension,” *J. Stat. Phys.* ***153*** *(2013), 377–399; arXiv:1306.5643*, 2013.
3. V. Popkov, A. Schadschneider, J. Schmidt, and G. M. Schütz, “Fibonacci family of dynamical universality classes,” *Proc. Natl. Acad. Sci. USA* ***112*** *(2015), no. 41, 12645–12650; arXiv:1505.04461*, 2015.
4. H. Spohn and G. Stoltz, “Nonlinear fluctuating hydrodynamics in one dimension: the case of two conserved fields,” *J. Stat. Phys.* ***160*** *(2015), 861–884; arXiv:1410.7896*, 2015.
5. V. Popkov, J. Schmidt, and G. M. Schütz, “Universality classes in two-component driven diffusive systems,” *J. Stat. Phys.* ***160*** *(2015), 835–860; arXiv:1410.8026*, 2015.

6. C. B. Mendl and H. Spohn, “Dynamic correlators of Fermi–Pasta–Ulam chains and nonlinear fluctuating hydrodynamics,” *Phys. Rev. Lett.* ***111*** *(2013), 230601; arXiv:1305.1209*, 2013.
7. H. van Beijeren, “Exact results for anomalous transport in one-dimensional Hamiltonian systems,” *Phys. Rev. Lett.* ***108*** *(2012), 180601; arXiv:1106.3298*, 2012.
8. S. G. Das, A. Dhar, K. Saito, C. B. Mendl, and H. Spohn, “Numerical test of hydrodynamic fluctuation theory in the Fermi–Pasta–Ulam chain,” *Phys. Rev. E* ***90*** *(2014), 012124; arXiv:1404.7081*, 2014.
9. H. Spohn, “Fluctuating hydrodynamics approach to equilibrium time correlations for anharmonic chains,” *in Thermal Transport in Low Dimensions, Lecture Notes in Phys.* ***921****, Springer, 2016, pp. 107–158; arXiv:1505.05987*, 2015.
10. T. Funaki and M. Hoshino, “A coupled KPZ equation, its two types of approximations and existence of global solutions,” *J. Funct. Anal.* ***273*** *(2017), 1165–1204; arXiv:1611.00498*, 2017.
11. A. Kupiainen and M. Marcozzi, “Renormalization of generalized KPZ equation,” *J. Stat. Phys.* ***166*** *(2017), 876–902; arXiv:1604.08712*, 2017.
12. I. Butelmann and G. R. M. Flores, “Scaling limit of stationary coupled Sasamoto–Spohn models,” *Electron. J. Probab.* ***27*** *(2022), paper no. 130; arXiv:2112.13810*, 2022.
13. Z. Chen, J. de Gier, I. Hiki, T. Sasamoto, and M. Usui, “Limiting current distribution for a two-species asymmetric exclusion process,” *Comm. Math. Phys.* ***395*** *(2022), no. 1, 59–142; arXiv:2104.00026*, 2022.
14. P. Nejjar, “KPZ statistics of second class particles in ASEP via mixing,” *Comm. Math. Phys.* ***378*** *(2020), no. 1, 601–623; arXiv:1911.09426*, 2020.
15. P. L. Ferrari and P. Nejjar, “Anomalous shock fluctuations in TASEP and last passage percolation models,” *Probab. Theory Related Fields* ***161*** *(2015), 61–109; arXiv:1306.3336*, 2015.
16. P. Nejjar, “GUE × GUE limit law at hard shocks in ASEP,” *Ann. Appl. Probab.* ***31*** *(2021), no. 1, 321–350; arXiv:1906.07711*, 2021.
17. A. Aggarwal, I. Corwin, and M. Hegde, “Scaling limit of the colored ASEP and stochastic six-vertex models,” *preprint (2024); arXiv:2403.01341*, 2024.
18. M. Balázs and T. Seppäläinen, “Order of current variance and diffusivity in the asymmetric simple exclusion process,” *Ann. of Math. (2)* ***171*** *(2010), 1237–1265; arXiv:math/0608400*, 2010.
19. P. A. Ferrari and J. B. Martin, “Stationary distributions of multi-type totally asymmetric exclusion processes,” *Ann. Probab.* ***35*** *(2007), no. 3, 807–832; arXiv:math/0501291*, 2007.
20. A. Aggarwal, M. Nicoletti, and L. Petrov, “Colored interacting particle systems on the ring: stationary measures from the Yang–Baxter equation,” *Compos. Math.* ***161*** *(2025), no. 8, 1855–1922; arXiv:2309.11865*, 2025.
21. A. Borodin and M. Wheeler, “Coloured stochastic vertex models and their spectral theory,” *Astérisque* ***437*** *(2022); arXiv:1808.01866*, 2022.

22. A. Aggarwal and A. Borodin, “Colored line ensembles for stochastic vertex models,” *preprint (2024); arXiv:2402.06868*, 2024.
23. P. A. Ferrari and C. Kipnis, “Second class particles in the rarefaction fan,” *Ann. Inst. H. Poincaré Probab. Statist.* ***31*** *(1995), no. 1, 143–154*, 1995.
24. F. Casini, C. Giardinà, and F. Redig, “Density fluctuations for the multi-species stirring process,” *J. Theoret. Probab.* ***37*** *(2024), no. 4, 3317–3354; arXiv:2307.05111*, 2024.
25. C. Franceschini, J. Kuan, and Z. Zhou, “Orthogonal polynomial duality and unitary symmetries of multi-species ASEP$(q, \theta)$ and higher-spin vertex models via $*$-bialgebra structure of higher rank quantum groups,” *Comm. Math. Phys.* ***405*** *(2024), no. 4, Paper No. 96; arXiv:2209.03531*, 2024.
26. G. Cannizzaro, P. Gonçalves, R. Misturini, and A. Occelli, “From ABC to KPZ,” *Probab. Theory Related Fields* ***191*** *(2025), 361–420; arXiv:2304.02344*, 2025.
27. J. Kuan and Z. Zhou, “Asymptotics of two-point correlations in the multi-species $q$-TAZRP,” *Braz. J. Probab. Stat.* ***38*** *(2024), no. 3, 339–346; arXiv:2305.00321*, 2024.
28. W. Groenevelt and C. Wagenaar, “A generalized dynamic asymmetric exclusion process: orthogonal dualities and degenerations,” *J. Phys. A: Math. Theor.* ***57*** *(2024), no. 37, 375202; arXiv:2306.12318*, 2024.
29. C. Wagenaar, “Dynamic generalizations of the asymmetric inclusion process, asymmetric Brownian energy process and their dualities,” *J. Stat. Phys.* ***193*** *(2026), no. 2, Art. 13; arXiv:2502.12709*, 2026.
30. M. Ayala and J. Zimmer, “Hydrodynamic limits and non-equilibrium fluctuations for the symmetric inclusion process with long jumps,” *Ann. Henri Poincaré (2025), online first; arXiv:2410.21933*, 2025.
31. H. D. Cunha and M. Sasada, “Stationary fluctuations for an exclusion process with mass and energy conservation,” *preprint (2026); arXiv:2606.01937*, 2026.
32. G. Cannizzaro, P. Cardoso, L. Gräfner, and A. Occelli, “Equilibrium fluctuations for a multi-species particle system with long jumps,” *preprint (2026); arXiv:2604.03777*, 2026.
33. F. Casini, R. Frassek, and C. Giardinà, “Duality for the multispecies stirring process with open boundaries,” *J. Phys. A: Math. Theor.* ***57*** *(2024), no. 29, 295001; arXiv:2312.15532*, 2024.
34. J. D. Nardis, D. Bernard, and B. Doyon, “Hydrodynamic diffusion in integrable systems,” *Phys. Rev. Lett.* ***121*** *(2018), 160603; arXiv:1807.02414*, 2018.
35. B. Doyon, “Lecture notes on generalised hydrodynamics,” *SciPost Phys. Lect. Notes* ***18*** *(2020); arXiv:1912.08496*, 2020.
36. B. Doyon, G. Perfetto, T. Sasamoto, and T. Yoshimura, “Ballistic macroscopic fluctuation theory,” *SciPost Phys.* ***15*** *(2023), no. 4, 136; arXiv:2206.14167*, 2023.
37. B. Derrida, M. R. Evans, V. Hakim, and V. Pasquier, “Exact solution of a 1D asymmetric exclusion model using a matrix formulation,” *J. Phys. A: Math. Gen.* ***26*** *(1993), 1493–1517*, 1993.

38. S. Corteel and L. K. Williams, “Tableaux combinatorics for the asymmetric exclusion process and Askey–Wilson polynomials,” *Duke Math. J.* ***159*** *(2011), no. 3, 385–415; arXiv:0910.1858*, 2011.
39. L. Cantini, J. de Gier, and M. Wheeler, “Matrix product formula for Macdonald polynomials,” *J. Phys. A: Math. Theor.* ***48*** *(2015), no. 38, 384001; arXiv:1505.00287*, 2015.
40. A. Kuniba, S. Maruyama, and M. Okado, “Multispecies TASEP and combinatorial $R$,” *J. Phys. A: Math. Theor.* ***48*** *(2015), no. 34, 34FT02; arXiv:1506.04490*, 2015.
41. J. B. Martin, “Stationary distributions of the multi-type ASEP,” *Electron. J. Probab.* ***25*** *(2020), paper no. 43; arXiv:1810.10650*, 2020.
42. D. Roy, A. Dhar, K. Khanin, M. Kulkarni, and H. Spohn, “Universality in coupled stochastic Burgers systems with degenerate flux Jacobian,” *J. Stat. Mech. (2024), 033209; arXiv:2401.06399*, 2024.
43. J. Schmidt, Ž. Krajnik, and V. Popkov, “Universality in driven systems with a multiply-degenerate umbilic point,” *J. Stat. Mech. (2026), 033202; arXiv:2601.04116*, 2026.
44. H. Weinberger, P. Comaron, and M. H. Szymańska, “Multicomponent Kardar–Parisi–Zhang universality in degenerate coupled condensates,” *Commun. Phys.* ***8*** *(2025), Art. 324; arXiv:2411.07095*, 2025.
45. J. Kuan, M. Landry, A. Lin, A. Park, and Z. Zhou, “Interacting particle systems with type D symmetry and duality,” *Houston J. Math.* ***48*** *(2022), no. 3, 499–538; arXiv:2011.13473*, 2022.
46. E. Rohr, K. S. Latha, and A. Yin, “A type D asymmetric simple exclusion process generated by an explicit central element of $\mathcal{U}_q(\mathfrak{so}_{10})$,” *Houston J. Math.* ***50*** *(2024), no. 2, 237–257; arXiv:2307.15660*, 2024.
47. E. Brodsky, E. R. Engel, C. Panish, and L. Stolberg, “Comparative analyses of the type D ASEP: stochastic fusion and crystal bases,” *arXiv:2407.21015*, 2024.
48. D. Blyschak, O. Burke, J. Kuan, D. Li, S. Ustilovsky, and Z. Zhou, “Orthogonal polynomial duality of a two–species asymmetric exclusion process,” *J. Stat. Phys.* ***190*** *(2023), Paper No. 101; arXiv:2209.11114*, 2023.
49. G. M. Schütz, “Duality relations for asymmetric exclusion processes,” *J. Stat. Phys.* ***86*** *(1997), 1265–1287*, 1997.
50. S. F. Edwards and D. R. Wilkinson, “The surface statistics of a granular aggregate,” *Proc. R. Soc. Lond. Ser. A* ***381*** *(1982), no. 1780, 17–31*, 1982.
51. M. Kardar, G. Parisi, and Y.-C. Zhang, “Dynamic scaling of growing interfaces,” *Phys. Rev. Lett.* ***56*** *(1986), 889–892*, 1986.
52. P. Gonçalves and M. Jara, “Nonlinear fluctuations of weakly asymmetric interacting particle systems,” *Arch. Ration. Mech. Anal.* ***212*** *(2014), 597–644; arXiv:1309.5120*, 2014.
53. M. Gubinelli and N. Perkowski, “Energy solutions of KPZ are unique,” *J. Amer. Math. Soc.* ***31*** *(2018), 427–471; arXiv:1508.07764*, 2018.

54. P. Gonçalves, M. Jara, and M. Simon, "Second order Boltzmann–Gibbs principle for polynomial functions and applications," *J. Stat. Phys.* ***166*** *(2017), 90–113; arXiv:1507.06076*, 2017.
55. M. Ayala, G. Carinci, and F. Redig, "Quantitative Boltzmann–Gibbs principles via orthogonal polynomial duality," *J. Stat. Phys.* ***171*** *(2018), 980–999; arXiv:1712.08492*, 2018.
56. M. Ayala, G. Carinci, and F. Redig, "Higher order fluctuation fields and orthogonal duality polynomials," *Electron. J. Probab.* ***26*** *(2021), 1–35; arXiv:2004.08412*, 2021.
57. A. Borodin, I. Corwin, and T. Sasamoto, "From duality to determinants for $q$–TASEP and ASEP," *Ann. Probab.* ***42*** *(2014), 2314–2382; arXiv:1207.5035*, 2014.
58. D. J. Aldous, "Stopping times and tightness," *Ann. Probab.* ***6*** *(1978), 335–340*, 1978.
59. I. Mitoma, "Tightness of probabilities on $C([0,1];\mathcal{S}')$ and $D([0,1];\mathcal{S}')$," *Ann. Probab.* ***11*** *(1983), 989–999*, 1983.
60. P. Massot, "leanblueprint: a plasTeX plugin allowing to write blueprints for Lean 4 projects," *software (version 0.0.20), https://github.com/PatrickMassot/leanblueprint*.
61. W. Groenevelt, "Orthogonal stochastic duality functions from Lie algebra representations," *J. Stat. Phys.* ***174*** *(2019), 97–119; arXiv:1709.05997*, 2019.
62. G. Carinci, C. Giardinà, F. Redig, and T. Sasamoto, "A generalized asymmetric exclusion process with $U_q(\mathfrak{sl}_2)$ stochastic duality," *Probab. Theory Related Fields* ***166*** *(2016), 887–933; arXiv:1407.3367*, 2016.
63. G. Carinci, C. Franceschini, and W. Groenevelt, "$q$-orthogonal dualities for asymmetric particle systems," *Electron. J. Probab.* ***26*** *(2021), 1–38; arXiv:2003.07837*, 2021.
64. G. F. Lawler and V. Limic, "Random Walk: A Modern Introduction," *Cambridge Stud. Adv. Math.* ***123****, Cambridge Univ. Press, 2010*.
65. E. A. Carlen, S. Kusuoka, and D. W. Stroock, "Upper bounds for symmetric Markov transition functions," *Ann. Inst. H. Poincaré Probab. Statist.* ***23*** *(1987), suppl., 245–287*, 1987.
66. V. V. Petrov, "Sums of Independent Random Variables," *Springer, 1975*.
67. N. H. Bingham, C. M. Goldie, and J. L. Teugels, "Regular Variation," *Encyclopedia of Mathematics and its Applications* ***27****, Cambridge Univ. Press, 1987*.
68. W. Feller, "An Introduction to Probability Theory and Its Applications," *Vol. II, 2nd ed., Wiley, 1971*.
69. G. M. Schütz, "Exact solution of the master equation for the asymmetric exclusion process," *J. Stat. Phys.* ***88*** *(1997), 427–445*, 1997.
70. C. A. Tracy and H. Widom, "Integral formulas for the asymmetric simple exclusion process," *Comm. Math. Phys.* ***279*** *(2008), 815–844; arXiv:0704.2633*, 2008.
71. C. Kipnis and C. Landim, "Scaling Limits of Interacting Particle Systems," *Grundlehren* ***320****, Springer, 1999*.

72. R. Holley and D. Stroock, “Generalized Ornstein–Uhlenbeck processes and infinite particle branching Brownian motions,” *Publ. Res. Inst. Math. Sci.* ***14*** *(1978), 741–788*, 1978.
73. P. Dittrich and J. Gärtner, “A central limit theorem for the weakly asymmetric simple exclusion process,” *Math. Nachr.* ***151*** *(1991), 75–93*, 1991.
74. R. Price, “A useful theorem for nonlinear devices having Gaussian inputs,” *IRE Trans. Inform. Theory* ***4*** *(1958), 69–72*, 1958.
75. W. F. Sheppard, “On the application of the theory of error to cases of normal distribution and normal correlation,” *Phil. Trans. R. Soc. A* ***192*** *(1899), 101–167*, 1899.
76. S. N. Ethier and T. G. Kurtz, “Markov Processes: Characterization and Convergence,” *Wiley, 1986*.